\documentclass[useAMS,usenatbib]{mn2e}

\usepackage{txfonts}
\usepackage{graphicx}
\usepackage{natbib}
\usepackage[latin1]{inputenc}
\usepackage{morefloats}

\title[A PAndAS view of M31 dwarf elliptical satellites]{A PAndAS view of M31 dwarf elliptical satellites: NGC147 and NGC185}
\author[D. Crnojevi\'c et al.]{D. Crnojevi\'c$^{1,2}$\thanks{Email:
denija.crnojevic@ttu.edu}, A. M. N. Ferguson$^{1}$, M. J. Irwin$^{3}$, A. W. McConnachie$^{4}$, E. J. Bernard$^{1}$, \newauthor M. A. Fardal$^{5}$, R. A. Ibata$^{6}$, G. F. Lewis$^{3,7}$, N. F. Martin$^{6}$, J. F. Navarro$^{8}$, N. E. D. No\"el$^{9}$, \newauthor S. Pasetto$^{10}$\\
$^{1}$Institute for Astronomy, University of Edinburgh, Royal Observatory, Blackford Hill, EH9 3HJ Edinburgh, UK\\
$^{2}$Physics Department, Texas Tech University, Lubbock, TX 79409, USA\\
$^{3}$Institute of Astronomy, University of Cambridge, Madingley Road, CB3 0HA Cambridge, UK\\
$^{4}$NRC Herzberg Institute of Astrophysics, 5071 West Saanich Road, Victoria, BC, V9E 2E7, Canada\\
$^{5}$Department of Astronomy, University of Massachusetts, Amherst, MA 01003, USA\\
$^{6}$Observatoire astronomique de Strasbourg, Universit\'e de Strasbourg, CNRS, UMR 7550, 11 rue de l'Universit\'e, F-67000 Strasbourg, France\\
$^{7}$Sydney Institute for Astronomy, School of Physics A28, The University of Sydney, NSW 2006, Australia\\
$^{8}$Department of Physics and Astronomy, University of Victoria, Victoria, BC, Canada\\
$^{9}$Department of Physics, University of Surrey, GU2 7XH Guildford Surrey, UK\\
$^{10}$University College London, Department of Space \& Climate Physics, Mullard Space Science Laboratory, Holmbury St. Mary, RH5 6NT Dorking Surrey, UK\\
}

\begin{document}

\date{Accepted 2014 September 24. Received 2014 September 21; in original form 2014 July 12}

\pagerange{\pageref{firstpage}--\pageref{lastpage}} \pubyear{}

\maketitle

\label{firstpage}

\begin{abstract}

  We exploit data from the Pan-Andromeda Archaeological Survey (PAndAS) to study
  the extended structures of  M31's dwarf elliptical companions,  NGC147 and NGC185. Our wide-field,
  homogeneous photometry allows to construct deep colour-magnitude diagrams (CMDs) which
  reach down to $\sim3$~mag below the red giant branch (RGB) tip. We trace the stellar components of
  the galaxies to surface brightness of $\mu_g \sim 32$~mag arcsec$^{-2}$ and show
  they have much larger extents ($\sim5$~kpc radii) than previously recognised. While
  NGC185 retains a regular shape in its peripheral regions, NGC147 exhibits pronounced
  isophotal twisting due to the emergence of symmetric tidal tails. 
  We fit single Sersic models to composite surface brightness profiles constructed
  from diffuse light and star counts and find that NGC147 has an effective radius almost 3 times that of
  NGC185. In both cases, the effective radii that we calculate are larger by a factor of $\sim2$
  compared to most literature values.  We also calculate revised total magnitudes of
  $M_g=-15.36\pm0.04$ for NGC185 and $M_g=-16.36\pm0.04$ for NGC147. Using photometric metallicities computed
  for RGB stars, we find NGC185 to exhibit a metallicity gradient of 
  [Fe/H]$\sim-0.15$~dex/kpc over the radial range 0.125 to 0.5~deg. On the other hand, NGC147 
  exhibits almost no metallicity gradient, $\sim-0.02$~dex/kpc from 0.2 to 0.6~deg. The
  differences in the structure and stellar populations in the outskirts of these systems suggest 
  that tidal influences have played an important role in governing the evolution of NGC147. 
 \end{abstract}


%________________________________________________________________

\begin{table*}
 \centering
 \caption{Assumed properties of NGC147 and NGC185. Columns are 
   (1): name of the galaxy; (2-3): equatorial coordinates
   from the NED database (http://ned.ipac.caltech.edu/); (4-5)
   Galactic longitude and latitude (NED); (6) absolute $V$ magnitude,
   obtained using the apparent, de-reddened magnitude listed in NED
   \citep[originally from][]{dev91} and the distance modulus derived by \citet{conn12}; 
(7-9): apparent TRGB $i$-band magnitude and deprojected distances from the Milky Way and 
from M31 \citep[][uncertainties stem from the resulting Posterior Probability Distributions]{conn12}; 
(10-11): mean extinction values (SDSS bands) from NED; (12-13): mean position angle 
(measured from north through east) and ellipticity as derived in this study (see Sect. \ref{analysis}).}
\label{tab1}
  \begin{tabular}{lccccccccccccc}
  \hline
  \hline
  Galaxy & $\alpha_{J2000}$ & $\delta_{J2000}$ & $l$ & $b$ & $M_V$ & $m_{i,TRGB}$ & $d_{MW}$ & $d_{M31}$ & $A_g$ & $A_i$ & $PA$ & $\epsilon$ \\
 & ($^h\, ^m\, ^s$)&($^\circ$ ' '')&(deg)&(deg)&(mag)&(mag)&(kpc)&(kpc)&(mag)&(mag)&($^\circ$)&\\
\hline

{NGC147}&00 33 12.12&48 30 31.5&119.82&-14.25&$-15.33$&$20.82^{+0.08}_{-0.08}$&$712^{+21}_{-19}$&118$^{+15}_{-15}$&0.570&0.293&$34.2\pm3.6$&$0.46\pm0.02$ \\
{NGC185}&00 38 57.97&48 20 14.6&120.79&-14.48&$-15.41$&$20.52^{+0.09}_{-0.08}$&$620^{+19}_{-18}$&181$^{+25}_{-20}$&0.608&0.313&$45.9\pm1.2$&$0.22\pm0.01$ \\

 \hline
 \hline
\end{tabular}
\end{table*}

\section{Introduction}

Dwarf galaxies have long been considered relatively simple objects due
to their small sizes. However, detailed studies of our Local Group (LG) dwarf
members have unveiled a great degree of complexity in their
resolved stellar populations and star formation histories (SFHs; e.g.,
\citealt{grebel97, mateo98, vandenb99}). Discrepancies between
their observed properties and theoretical predictions have often 
been used to argue against the
widely-accepted $\Lambda$~CDM model, although uncertainties in modelling
baryonic physics may also play a role \citep[for a recent discussion
see e.g.,][]{weinberg13}.  Nonetheless, dwarf galaxies are exciting testbeds for
refining our view of galaxy formation and evolution, and represent a
vital step in the understanding of more massive and complex objects.

As a result of their small masses, dwarf galaxies are particularly
sensitive to processes that affect gas content, and thus star
formation, and both internal feedback and environmental effects are
likely to play major roles in governing their evolution. For example,
the origin of dwarf elliptical (dE) galaxies is a longstanding
question: are they simply scaled-down versions of more massive
early-type galaxies, which experience the main episode of star
formation at early epochs, or do they result from the morphological
transformation of gas-rich dwarf irregular (dIrr) galaxies which are
depleted of their gas content through environmental processes (e.g.,
ram-pressure stripping, tidal interactions)?  The role of internal
versus environmental factors has been discussed extensively in the
literature \citep[see, e.g.,][]{mateo98, grebel03} but there is
as of yet no consensus.  Investigating the structural profiles and
stellar populations gradients in these systems has the potential to
help distinguish between possible origins.

It is within the LG that the resolved stellar populations of dwarf
galaxies can be studied with the greatest precision.  Only four dE
galaxies are known within the LG (NGC185, NGC147, NGC205 and M32)
and all of these are likely satellites of M31.  Lying in M31's remote
halo, NGC185 and NGC147 were first recognised as LG members by
\citet{baade44}.   Accurate distances, derived with the tip of the red
giant branch (TRGB) method, place NGC185 at $\sim620$ and NGC 147 
at $\sim710$~kpc from us\footnote{At these distances, 1~kpc$=$5.6~arcmin 
(NGC185) and 1~kpc$=$4.8~arcmin (NGC147).}, and at $\sim180$ and 
$\sim120$~kpc from the centre of M31 \citep[][see Tab.  \ref{tab1}]{conn12}. 
Their 3D separation  is only $93\pm27$~kpc and they have heliocentric radial velocities of 
$\sim-203$~km/s and $\sim-193$~km/s, respectively
\citep{geha10}. There has been much debate about whether they form a physical
system or not \citep[e.g.][]{vandenbergh98,watkins13, fattahi13, evslin13}.  Despite
having similar luminosities,
the two dEs show some marked differences in their internal properties.
NGC 185 hosts a population of young stars, some of which formed as
recently as $\sim100$~Myr ago \citep[e.g.][]{martinez99}. It also has
central dust lanes and $\sim3.0\times10^5$~M$_\odot$ of neutral gas, an
amount that is consistent with stellar mass loss originating from the
most recent star formation event \citep{marleau10}.  On the other
hand, NGC147 does not contain significant dust or gas, and shows no
evidence for a population younger than $\sim1$~Gyr
\citep[e.g.][]{sage98, marleau10}. The mean metallicities and metallicity 
gradients have been
investigated in both systems with varying conclusions, possibly
reflecting the different analysis methods and areal coverages of the
individual studies \citep[e.g.][]{han97, martinez99, battinelli04a,
  battinelli04b, goncalves07, geha10, goncalves12, ho14}.   A recent
spectroscopic study of NGC185 and NGC147 probed stellar kinematics to
large radii and discovered significant rotation in both galaxies
\citep{geha10}, contrary to what had been previously found by smaller
FoV studies \citep{bender91, simien02}.  This provides important
insight into the formation mechanism of this class of galaxies,
suggesting a rotationally-supported origin possibly followed by a
morphological transformation.
  
In this work, we present a photometric analysis of the stellar content
of NGC147 and NGC185 derived from very wide-field survey data.  
Many galaxies have recently been shown to have large  
low surface brightness extensions 
\citep[e.g.][]{barker12, bernard12b, crnojevic13, ibata14}, the properties of the
which hold important clues to their past
histories, and it is natural to ask whether these two systems do too. We exploit 
data from the Pan-Andromeda Archaeological Survey (PAndAS)
that imaged the M31 halo out to projected distances of $\sim150$~kpc
\citep{mcconnachie09} and which provides deep uniform photometry out
to unprecedented radii for both dwarfs. This dataset has already led
to the discovery of a prominent stellar stream emanating from NGC147,
which is reported in detail elsewhere \citep[e.g.,][Irwin et al., in
prep.]{ibata14, martin13}, as well as several new globular clusters
around both systems \citep{veljanoski13}.  In this paper, we analyse
the resolved stellar populations across large swaths of the two dEs
and re-examine their extents, structural properties and chemical
contents.

The paper is organized as follows: in \S \ref{obs} we present the data
and the photometry, and show the resulting colour-magnitude diagrams
(CMDs) in \S \ref{cmds}.  We analyse the spatial distribution of
populations in \S \ref{analysis}, their radial density and surface
brightness profiles in \S \ref{prof_sec}, while in \S \ref{mdfss} we
derive photometric metallicity distribution functions (MDFs) and
investigate the metallicity gradients. We discuss our results in \S
\ref{disc}, and draw our conclusions in \S \ref{concl}.

%________________________________________________________________

\begin{figure*}
  \centering
 \includegraphics[width=8.5cm]{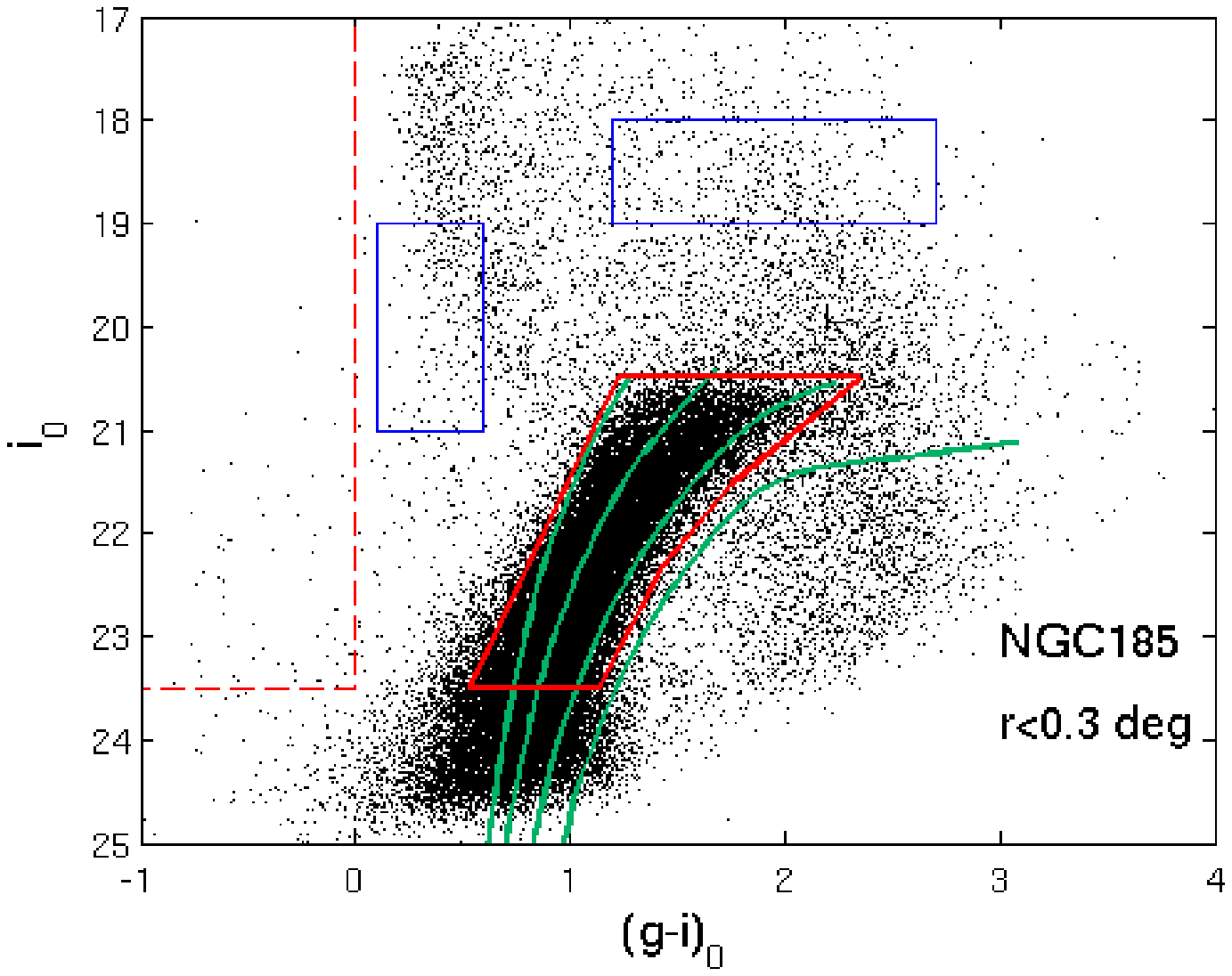}
 \includegraphics[width=8.5cm]{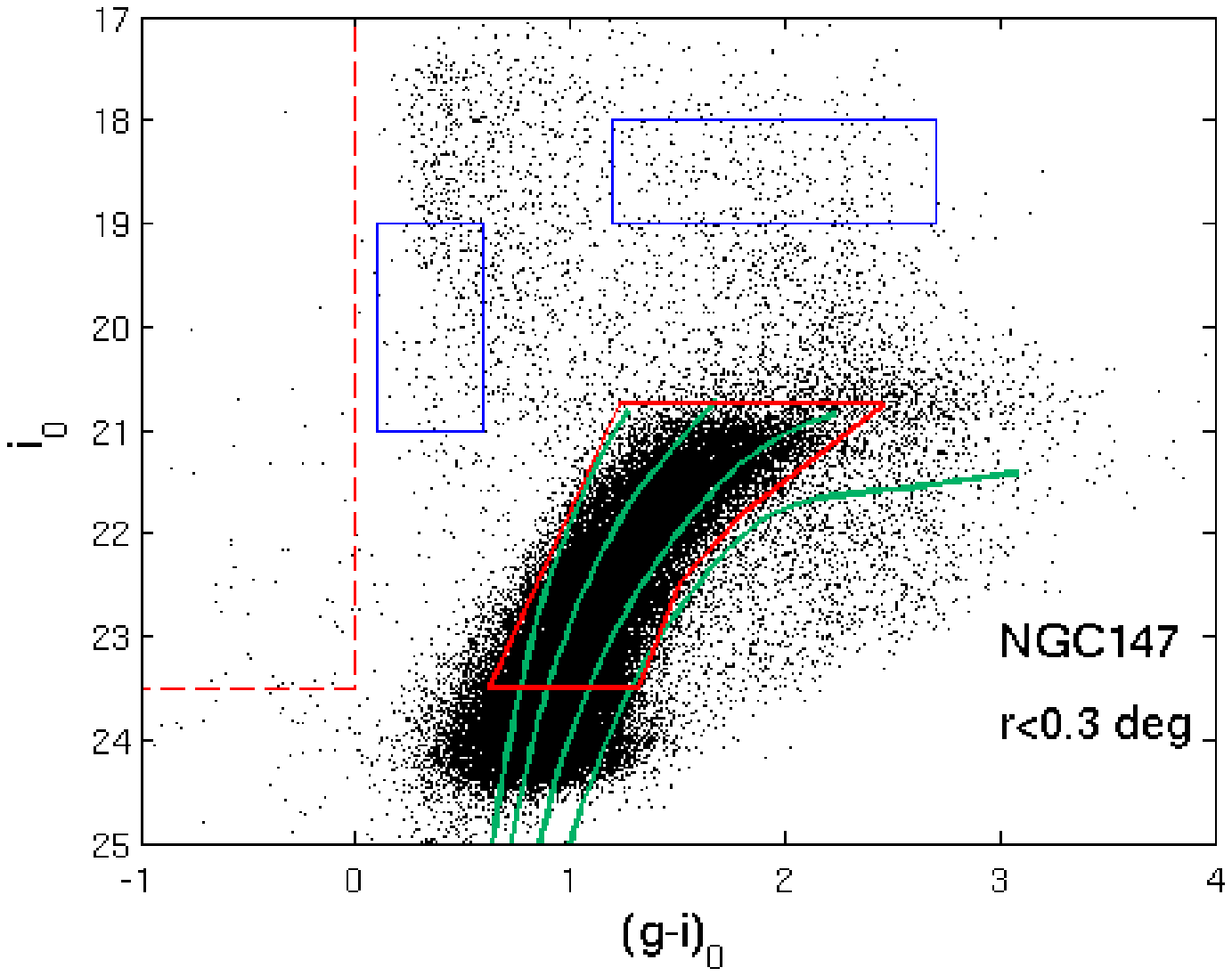}
 \caption{De-reddened CMDs of the dEs, showing only stars falling
   within an ellipse of semi-major axis 0.3~deg.  Overlaid are
   Dartmouth isochrones (green lines) with a fixed age of 12~Gyr and
   varying metallicity ([Fe/H]$=-2.5$, $-1.3$, $-0.7$ and $-0.3$),
   shifted to the appropriate distance. The adopted RGB boxes are
   outlined in red, while blue boxes indicate areas dominated by 
   foreground sequences (Galactic halo: $0.1<(g-i)_0<0.6$,
   $19<i_0<21$; Galactic disc: $1.2<(g-i)_0<2.7$, $18<i_0<19$). The
   dash-dotted red line isolates candidate young stars. We do not
   consider luminous AGB stars in this work as their study is deferred
   to a future paper incorporating a NIR dataset.}
\label{cmd_glob}
\end{figure*}

\section{Observations and photometry} \label{obs}

The PAndAS survey was conducted with the MegaPrime/MegaCam camera on
the 3.6~m Canada-France-Hawaii Telescope (CFHT) under excellent
average seeing conditions ($<0.8''$). MegaCam consists of a mosaic of
36 $2048\times4612$ CCDs with a total FoV of $\sim1\times1$~deg$^2$
and pixel scale $0.187''$ per pixel.  The effective area covered by
each pointing is about $0.96\times0.94$~deg$^2$ due to small gaps
between the chips. In total, roughly 400~deg$^2$ was mapped around M31
and M33 via 413 distinct MegaCam pointings.  The observing strategy,
data processing and calibration have been fully described elsewhere
\citep{mcconnachie09, mcconnachie10, richardson11, ibata14}.

As a brief summary, the raw data are pre-processed through the
\emph{Elixir} CFHT pipeline, and the source detection and aperture
photometry are obtained with the CFHT/MegaPrime adapted version of the
Cambridge Astronomical Survey Unit (CASU) pipeline \citep{irwin01}.
PSF-fitting photometry is subsequently performed using DAOPHOT/ALLSTAR
\citep{stetson87}.  Although PSF-fitting photometry leads to smaller
photometric uncertainties \citep[see][]{ibata14}, the quality of fit
parameters vary greatly between CCDs and thus cannot be used to clean
our catalogue of non-stellar sources.  Instead, the PSF photometry and
the CASU aperture photometry are cross-matched and the source
classifications from the latter are adopted.  For the bulk of our
analysis, only objects classified as stellar or probably stellar (i.
e. within $2\sigma$ of the stellar locus; \citealt{irwin01}) in both
bands are used.  The resulting $g$ and $i$ magnitudes are on the AB
system, and the final band-merged catalogue provides both
morphological classification and accurate astrometry for each source.
Sources in our catalogue have a signal-to-noise ratio of $\sim5$
(corresponding to a photometric error of 0.2~mag) at $i_0\sim24-24.5$
and $g_0\sim25-25.5$.

Throughout this study we use de-reddened magnitudes, adopting the
extinction values obtained from the \citet{schlegel98} $E(B-V)$ maps
interpolated on a star-by-star basis. Additionally, we apply the
\citet{bonifacio00} correction for $E(B-V)>0.1$ (their formula 1),
which returns slightly smaller values than the original ones.  The
mean extinction for a $\sim3\times3$~deg$^2$ field centered on NGC185
and NGC147 is $<E(B-V)>\sim0.13$ and presents spatial variations of up
to $\sim40\%$. We apply the following conversion to obtain the final
de-reddened magnitudes: $g_0 = g-3.793~E(B-V)$ and $i_0 =
i-2.086~E(B-V)$ \citep{schlegel98}.

Tab. \ref{tab1} lists the properties of the two dwarf galaxies. The distance 
to the dEs is taken from \citet{conn12}, which has been
derived from the TRGB luminosity via a Bayesian approach using the
same PAndAS dataset.  We have recomputed the
position angle $PA$, ellipticity $\epsilon$ and effective radius
$r_{\rm eff}$ for each galaxy, given the larger extent of our data
with respect to previous surveys (details in Sect. \ref{analysis}).
Throughout this paper, we use standard coordinates (centered on either
NGC185 or NGC147) and refer to galactocentric projected elliptical
radii ($r=\sqrt{x^2+y^2/(1-\epsilon)^2}$) computed using our mean
values of position angle ($PA$) and $\epsilon$. 
%________________________________________________________________

\section{Colour-magnitude diagrams} \label{cmds}

\begin{figure*}
  \centering
 \includegraphics[width=10cm]{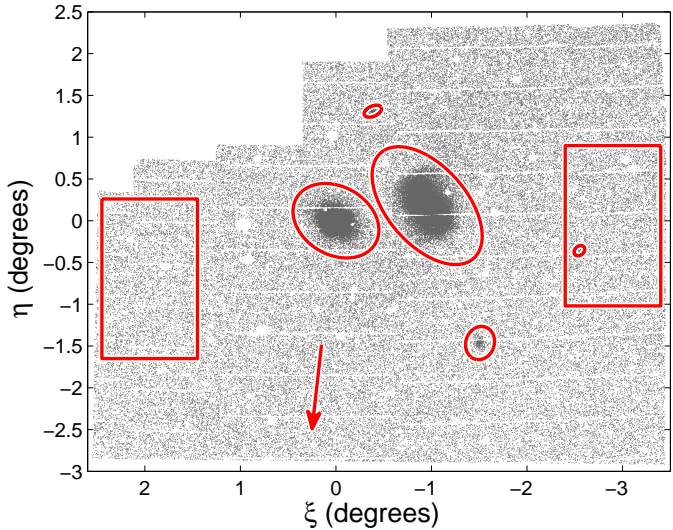}
 \caption{Spatial distribution of sources falling within the NGC185
   RGB box (in standard coordinates centered on NGC185). North is up
   and east is left. The red boxes indicate the position of the two
   adopted field regions (each covering a $\sim1.9$~deg$^2$ area) used
   to monitor contaminants.  Large red ellipses are drawn with radii
   of 0.5~deg around NGC185 (left) and 0.8~deg around NGC147 (right).
   Also shown are the dwarf spheroidal galaxies Cass II ($\xi=-0.39
   ^\circ$, $\eta=1.31 ^\circ$, And XXV ($\xi=-1.51 ^\circ$,
   $\eta=-1.46 ^\circ$) and And XXVI ($\xi=-2.55 ^\circ$, $\eta=-0.36
   ^\circ$), where the ellipse in each case indicates $4 r_{\rm eff}$.
   The arrow points in the direction of M31. } \label{field}
\end{figure*} 

\begin{figure*} \centering
  \includegraphics[width=17cm]{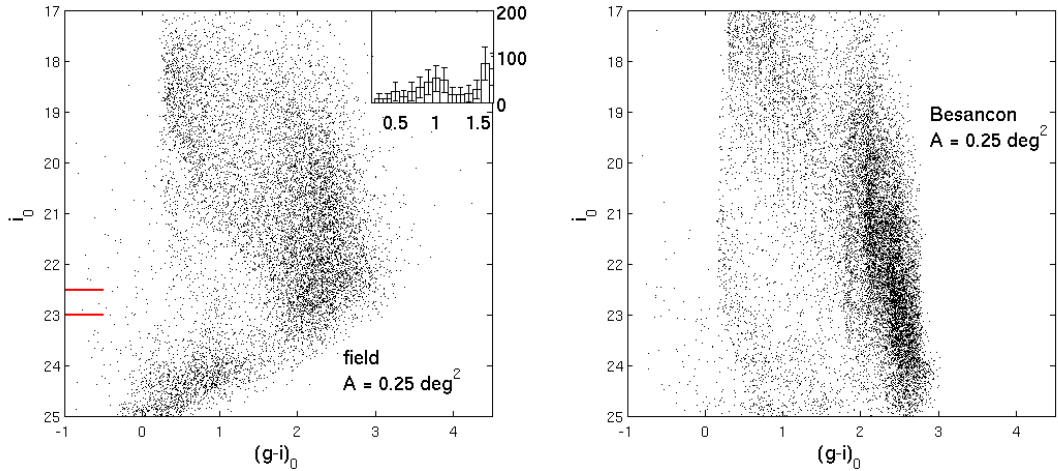} 
\caption{The CMDs
    of contaminant sources. \emph{Left panel}: the CMD of point
    sources lying in the field selection boxes, scaled to an arbitrary
    area of 0.25~deg$^2$. The inset shows the colour histogram of
    sources per unit area with
    $(i_{0,TRGB\_NGC185}+2.0)<i_0<(i_{0,TRGB\_NGC185}+2.5)$ (this
    magnitude range is denoted with red lines on the y-axis). At
    fainter magnitudes, unresolved galaxies dominate the CMD.
    \emph{Right panel}: simulated Galactic foreground from the
    Besan\c{c}on model, scaled to an area of 0.25~deg$^2$. The
    simulation has been convolved with photometric errors (as a
    function of magnitude only) stemming from our photometry, but is
    not corrected for the data incompleteness.}  \label{cmd_field}
\end{figure*}

In Fig. \ref{cmd_glob} we show de-reddened
colour-magnitude diagrams (CMDs) for NGC185 and NGC147. We plot all
stars in each system lying within an ellipse of semi-major axis
$0.3$~deg to highlight the main CMD sequences.  The most prominent
feature in the CMDs is a broad red giant branch (RGB), indicative of
old ($\sim2$ to 13~Gyr) populations that have evolved off the main
sequence (MS).  Dartmouth isochrones in the CFHT photometric system
\citep{dotter08, mcconnachie10} have been overlaid on the RGB. The
isochrones have been shifted to the distance of each galaxy and have
an age of 12~Gyr, [$\alpha$/Fe] $=+0.0$ and varying metallicity to
encompass the full width of the RGB.  For our subsequent analysis, we
define a selection box on the main RGB locus with the bright limit set
by the TRGB luminosity and a faint limit of $i_0=23.5$.  The faint
limit has been chosen to limit incompleteness in our data and minimise
contamination from unresolved galaxies which become significant at
magnitudes fainter than this (see \citet{ibata14} and Fig.
\ref{cmd_field}).  

The other main features visible in the CMDs are due
to foreground stars which are especially prominent given the low
Galactic latitude of the two systems.  The almost vertical sequence at
$0.2\lesssim(g-i)_0\lesssim1.0$ and $g_0\lesssim21$ results from halo
turnoff stars, while the broad diagonal sequence with
$1.5\lesssim(g-i)_0\lesssim3.0$ is due to disc dwarfs. Short-lived, luminous asymptotic giant branch (AGB)
stars are expected to be found above the TRGB and are indicative of
intermediate-age populations ($\sim0.5-8$~Gyr).  Such populations are
indeed visible up to $\sim1$~mag above the TRGB and redward of the RGB
box, at colours $1.5\lesssim(g-i)_0\lesssim3$, especially for NGC147.
However, this region of the CMD is significantly contaminated by
foreground stars in optical bands, such that an accurate analysis of
AGB stars is not feasible from the PAndAS data alone. We defer the
study of intermediate-age populations in the target dEs to a future
paper which will also incorporate near-infrared (NIR) data, enabling a
cleaner and more robust separation from foreground stars.  

Finally, a
few objects are visible at colours bluer than $(g-i)_0\lesssim0$
(vertical dashed line in Fig. \ref{cmd_glob}), suggesting the presence
of young, massive MS stars.  While NGC185 has been previously shown to
have signs of recent star formation in
its main body \citep{lee93b, martinez99, butler05, marleau10}, NGC147
is not known to contain young stars.  Some foreground stars could
contaminate this CMD region, however. We define a selection box for
blue stars at $-1.0<(g-i)_0<0.0$, $17<i_0<23.5$ and consider in
detail their spatial distribution in the following analysis, though we
note that crowding in the central regions -- where these sources 
typically lie -- limits straightforward interpretation.  

\subsection{Foreground/background contamination} 

While
our photometric catalogue will be dominated by stars in NGC147 and
NGC185, there will also be contaminant sources.  Accounting for these
is crucially important when analysing the low surface density
peripheral regions of the two dwarfs.  The are three types of
contaminating sources to consider -- Galactic foreground stars, M31
halo stars, and unresolved background galaxies. While M31 halo stars
and background galaxies are expected to be relatively smoothly
distributed over our FoV (we are considering large galactocentric
distances from M31 where the halo density falls off slowly,
\citealt{ibata14}), the contribution from Galactic disc stars
significantly increases in the north-west direction due to the low
Galactic latitude ($b\sim -14^\circ$) of the systems. 

 In order to
account for contaminants, we select two large rectangular ``field''
regions that lie east and west of the dEs (see Fig. \ref{field}), each
with an area of $\sim1.9$~deg$^2$.  These large areas minimise
sensitivity to any possible localized substructure in M31's halo.  The
east region contains $\sim75000$~stars, while the west region contains
$\sim85000$~stars; this difference is in the right sense to be
consistent with gradient in foreground density due to the Galactic
disc.  Our field regions are chosen to avoid the north-south direction
because of the presence of NGC147's extended tidal tails and the dwarf
spheroidal galaxies Cass II (Irwin et al., in prep.)  and And XXV
\citep{richardson11}.  A third
low-mass galaxy, And XXVI (the least luminous of the three), lies
within our west field region but we have excised stars falling within
a 4~$r_{\rm eff}$ ellipse centered on this dwarf (there are virtually
no stars belonging to And XXVI outside this radius).  

The CMD
of the combined field regions is shown in the left panel of Fig.
\ref{cmd_field}, scaled to an arbitrary area of 0.25~deg$^2$ to ease
the identification of the main features.  This is comparable to the
surveyed area within $r=0.3$~deg (after gap subtraction) for NGC185
($\sim0.20$~deg$^2$) and NGC147 ($\sim0.27$~deg$^2$) and hence should
represent the level of contamination present in the CMDs of Fig. \ref{cmd_glob}.
Besides the prominent foreground sequences, sources found at
magnitudes greater than $i_0\sim23.5$ (i.e. below our RGB box faint
limit) are largely unresolved background galaxies \citep[see, e.g., Fig 9 of][]{barker12}. 
The RGB of M31's stellar halo is barely visible due
to its low density at these galactocentric radii.  Finally, a few
sources bluer than $(g-i)_0\sim0$ are also present, representing an
important source of contamination for the identification of possible
young stars belonging to the dEs.  

The right panel of Fig.
\ref{cmd_field} shows the Galactic foreground in the direction of
NGC185 and NGC147 as predicted by the Besan\c{c}on model \citep{robin03}. 
Stars have been simulated over an area of 2.5~deg$^2$ centered
at an ({\it{l,b}}) that lies between the two dwarfs and subsequently
convolved with the observational errors from our photometry (as a
function of magnitude only).  We have not applied a correction for
incompleteness.  From the final catalogue of foreground contaminants,
we extract a subset of stars by scaling their numbers to an area of
0.25~deg$^2$ in order to match the left panel of Fig. \ref{cmd_field}.
The simulated foreground sequences are broadly in agreement with the
observed ones.  We consider stars from the observed field CMD within
the same RGB selection boxes drawn in Fig.  \ref{cmd_glob} as our
contaminant population in subsequent analyses.  Finally, for the
observed field CMD, we compute the colour histogram of sources that
lie within the RGB selection box and have magnitudes in the range
$(i_{0,TRGB\_NGC185}+2.0)<i_0<(i_{0,TRGB\_NGC185}+2.5)$.  The
overdensity of sources at colours $(g-i)_0\sim1$ results from M31 halo
stars, while at redder colours the Galactic foreground dominates.

\begin{figure*}
  \centering
 \includegraphics[width=13cm]{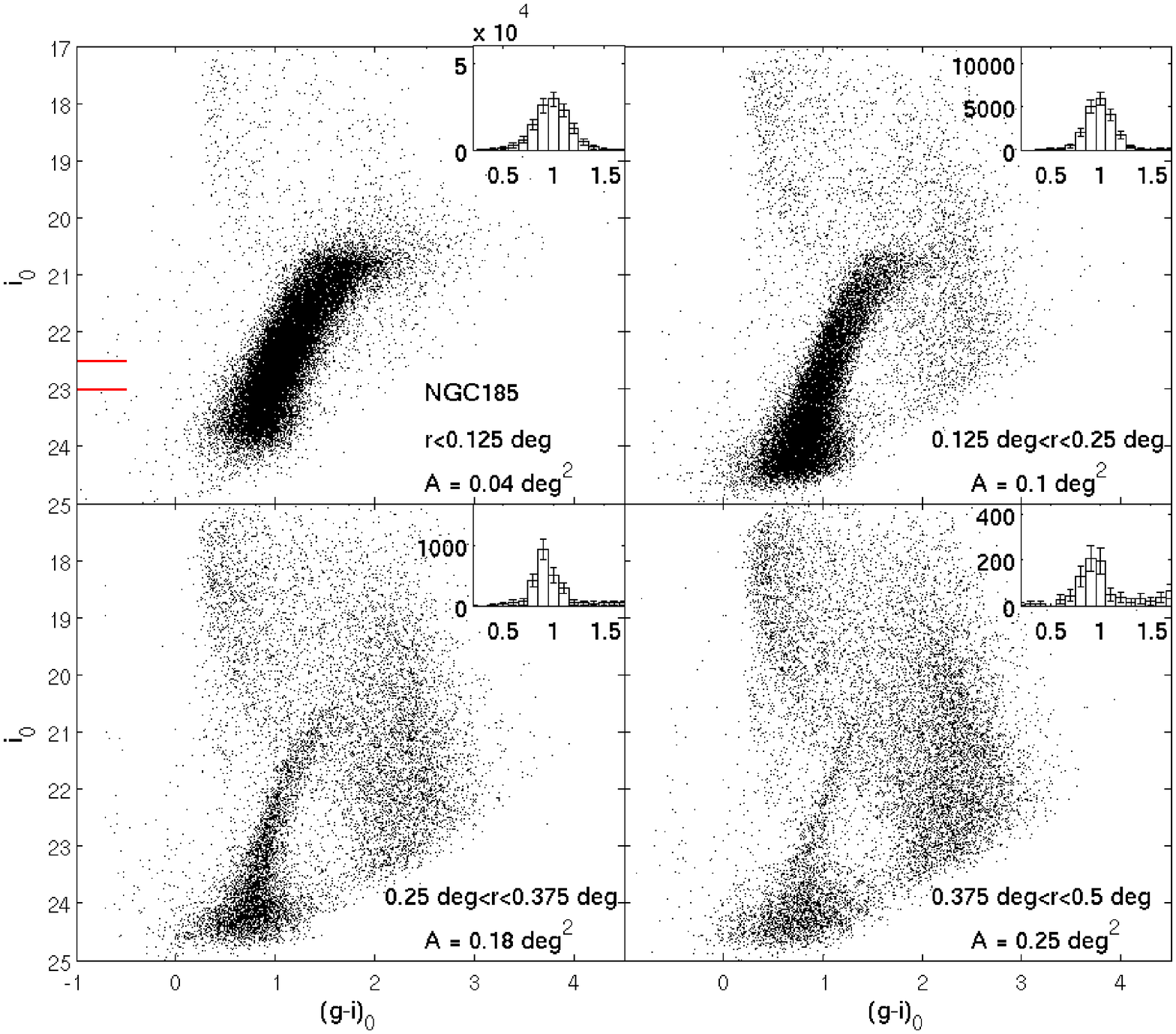}
 \includegraphics[width=13cm]{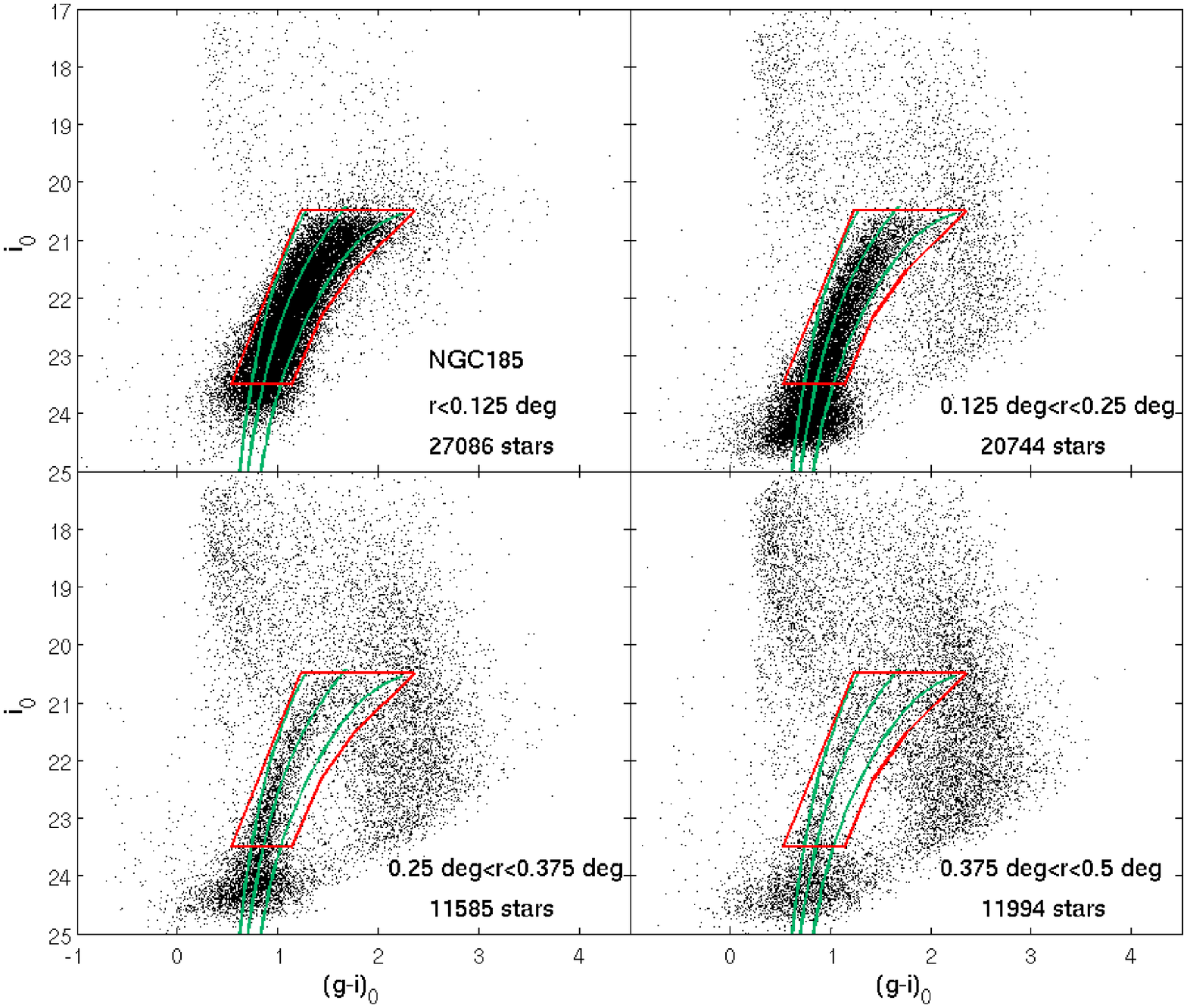}
 \caption{CMDs in four radial bins (in 0.125~deg steps) for NGC185,
   assuming the average $PA$ and ellipticity derived in this study.
   \emph{Upper panels}: The radial limits and enclosed areas are
   reported in each subpanel. The insets show colour histograms of
   sources per unit area with
   $(i_{0,TRGB\_NGC185}+2.0)<i_0<(i_{0,TRGB\_NGC185}+2.5)$ (this
   magnitude range is denoted with red lines on the y-axis).
   \emph{Lower panels}: As above, but with Dartmouth isochrones of
   fixed age (12~Gyr) and varying metallicity ([Fe/H]$=-2.5$, $-1.3$
   and $-0.7$) overlaid. Also shown is the adopted RGB selection box
   and the number of stars per bin.}
\label{cmd_spat_185}
\end{figure*}

\begin{figure*}
  \centering
 \includegraphics[width=13cm]{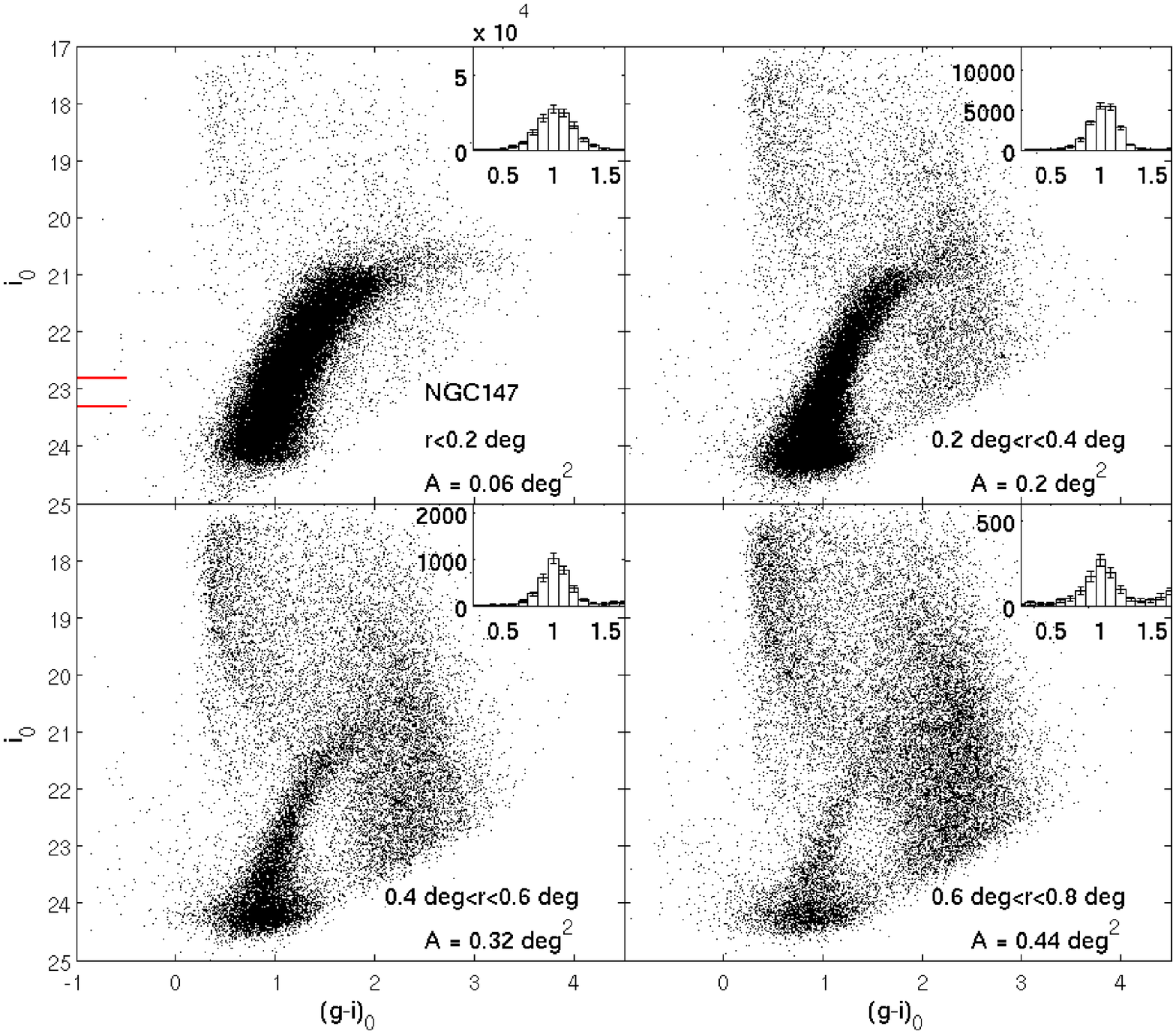}
 \includegraphics[width=13cm]{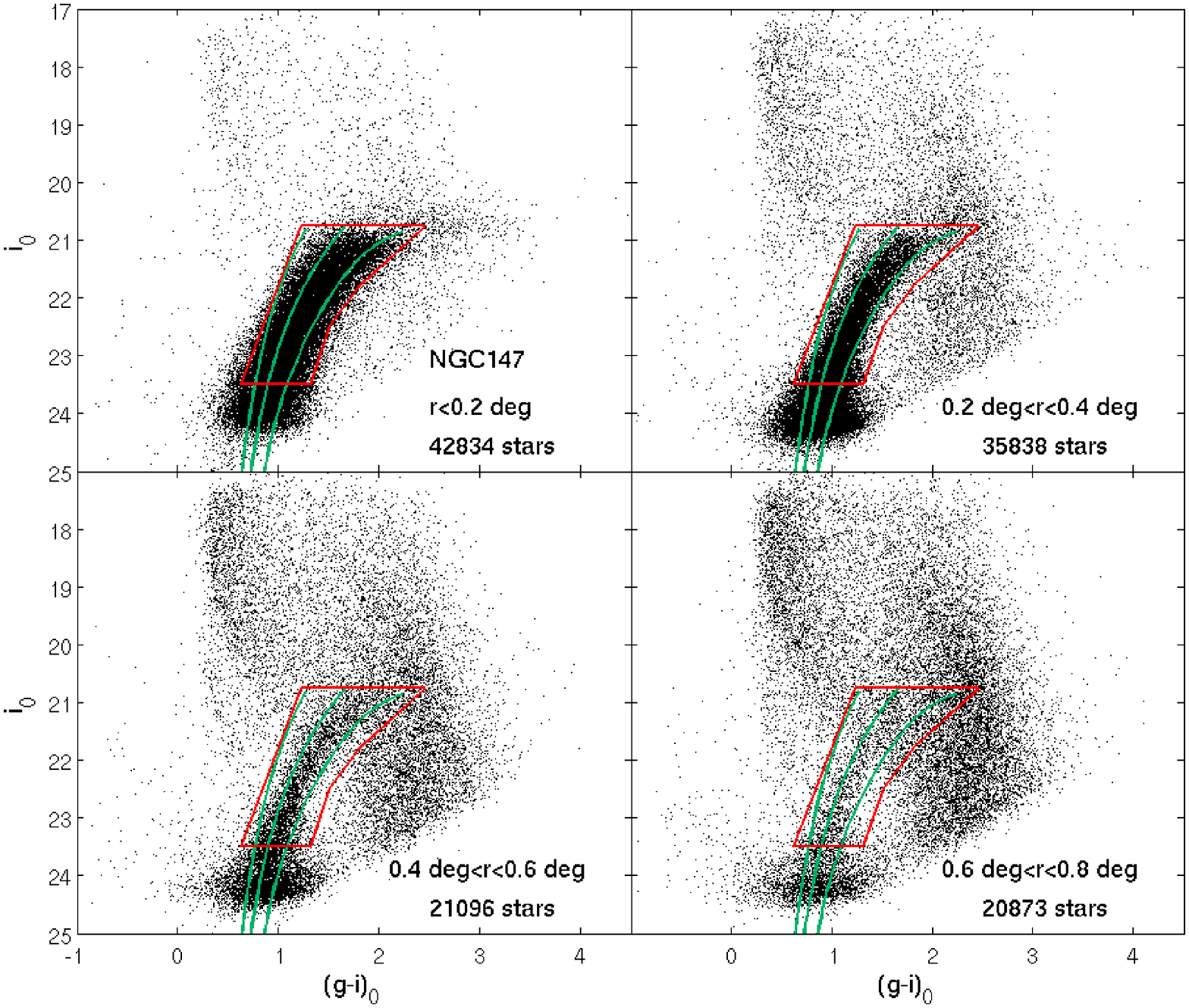}
 \caption{Same as Fig. \ref{cmd_spat_185}, for NGC147 and with radial bins of 0.2~deg width. 
   The insets show colour histograms of sources per unit area with 
   $(i_{0,TRGB\_NGC147}+2.0)<i_0<(i_{0,TRGB\_NGC147}+2.5)$
  (this magnitude range is denoted with red lines on the y-axis).}
\label{cmd_spat_147}
\end{figure*}

\subsection{Colour-magnitude diagrams as a function of radius}

To explore how the stellar populations vary across the face of
each galaxy, we construct CMDs in four elliptical bins spanning
a range of radii and assuming the average $PA$s and ellipticities
derived in this study (see Sect. \ref{analysis}).
For NGC185, we restrict our analysis here and throughout
the rest of the paper to $\le0.5$~deg
to avoid contamination from NGC147's tidal tails.
The CMDs are shown in Fig. \ref{cmd_spat_185} and 
\ref{cmd_spat_147}, with and without isochrones overlaid. 

In both systems, a clear signature of an RGB population is seen at all
radii demonstrating the significant extents of these dwarf galaxies.
In addition, the width of the RGB clearly decreases as a function of
increasing galactocentric radius, suggestive of a smaller dispersion
in metallicity in the outermost regions.  To demonstrate this, colour
histograms for RGB stars (per unit area) in a narrow range of
magnitudes ($(i_{0,TRGB}+2.0)<i_0<(i_{0,TRGB}+2.5)$), assuming the
TRGB of each galaxy) are constructed and shown as insets in Fig.
\ref{cmd_spat_185} and \ref{cmd_spat_147}.  For NGC185, the width of
the histogram distribution decreases from the first to the fourth
radial bin, and the peak of the distribution also shifts bluewards.
For NGC147, the width decreases from the
first two to the second two radial bins, but the peak of the
distribution does not significantly change. We note that the innermost
radial bin for each galaxy is more incomplete at faint magnitudes, due
to stellar crowding in these regions.

%________________________________________________________________

\begin{figure*}
  \centering
 \includegraphics[width=8.5cm]{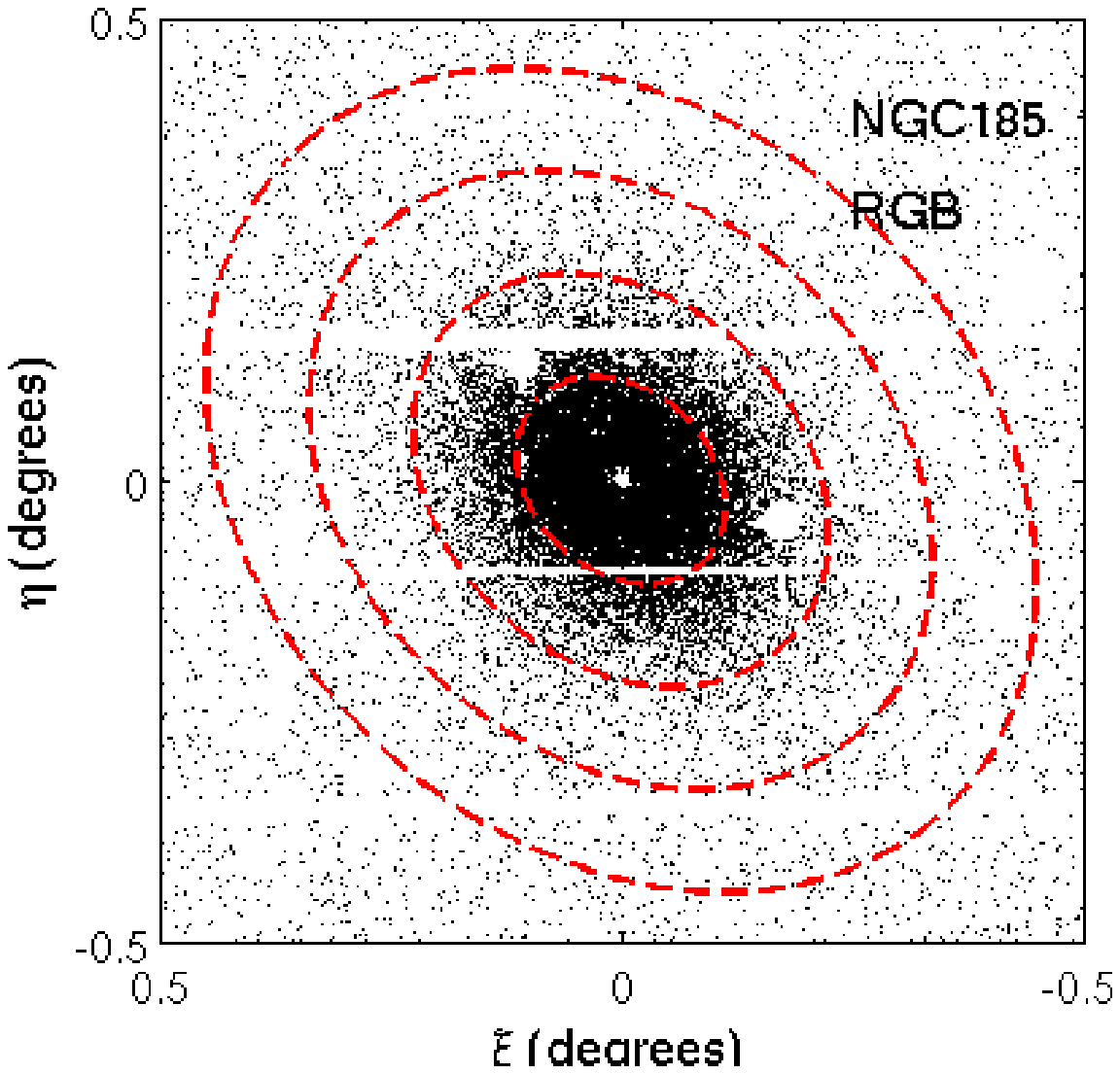}
 \includegraphics[width=8.5cm]{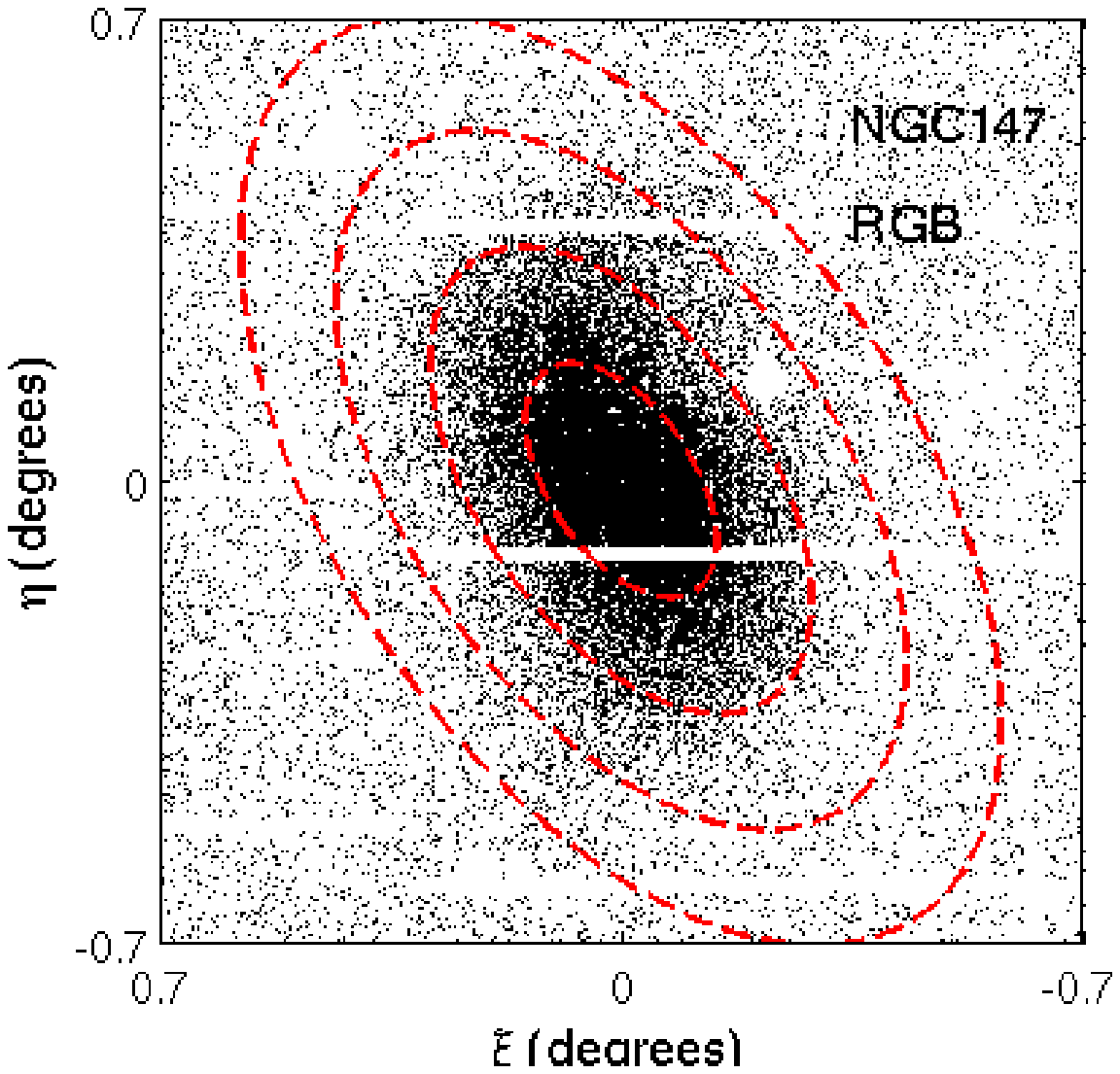}
\includegraphics[width=8.5cm]{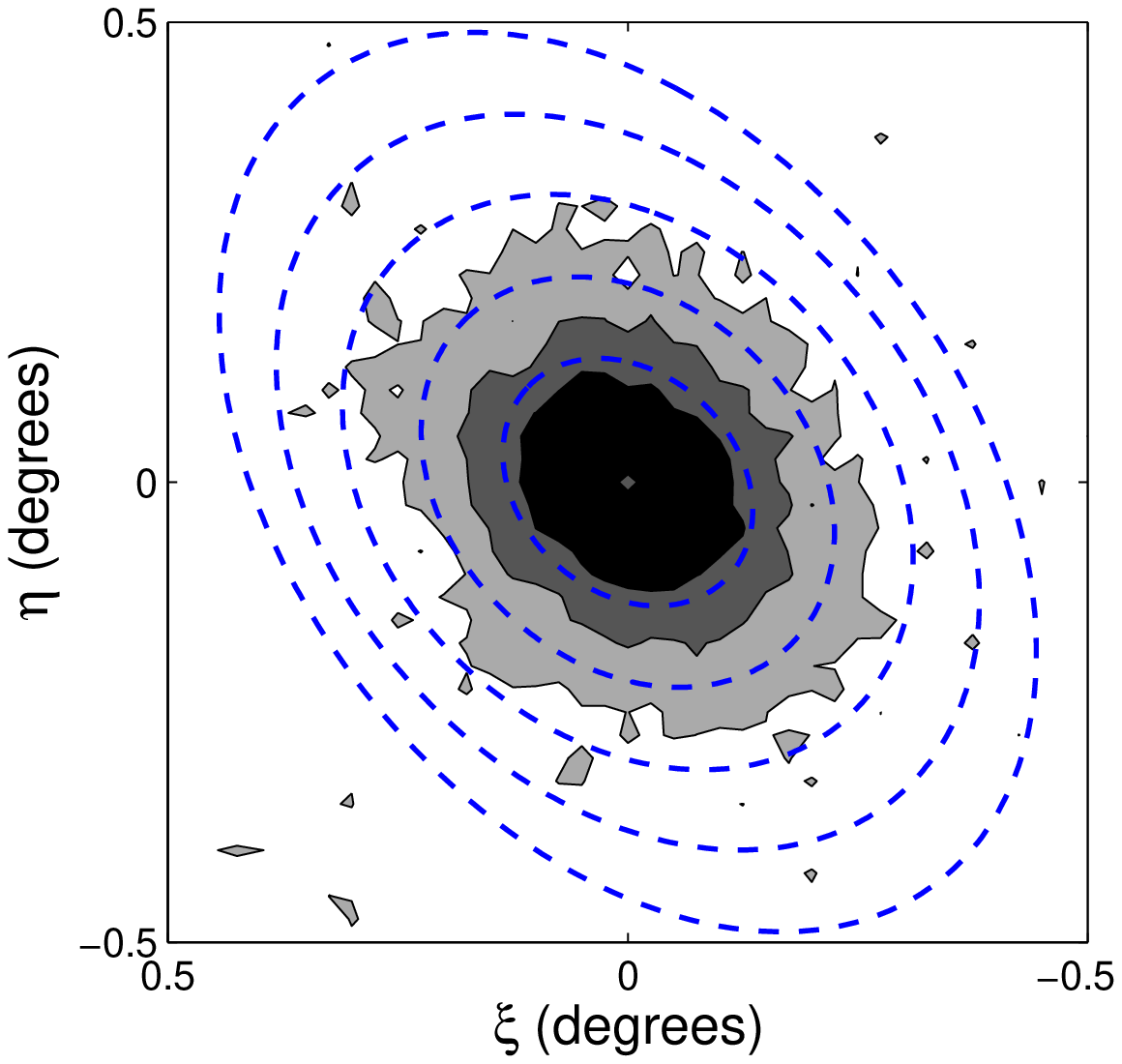}
 \includegraphics[width=8.5cm]{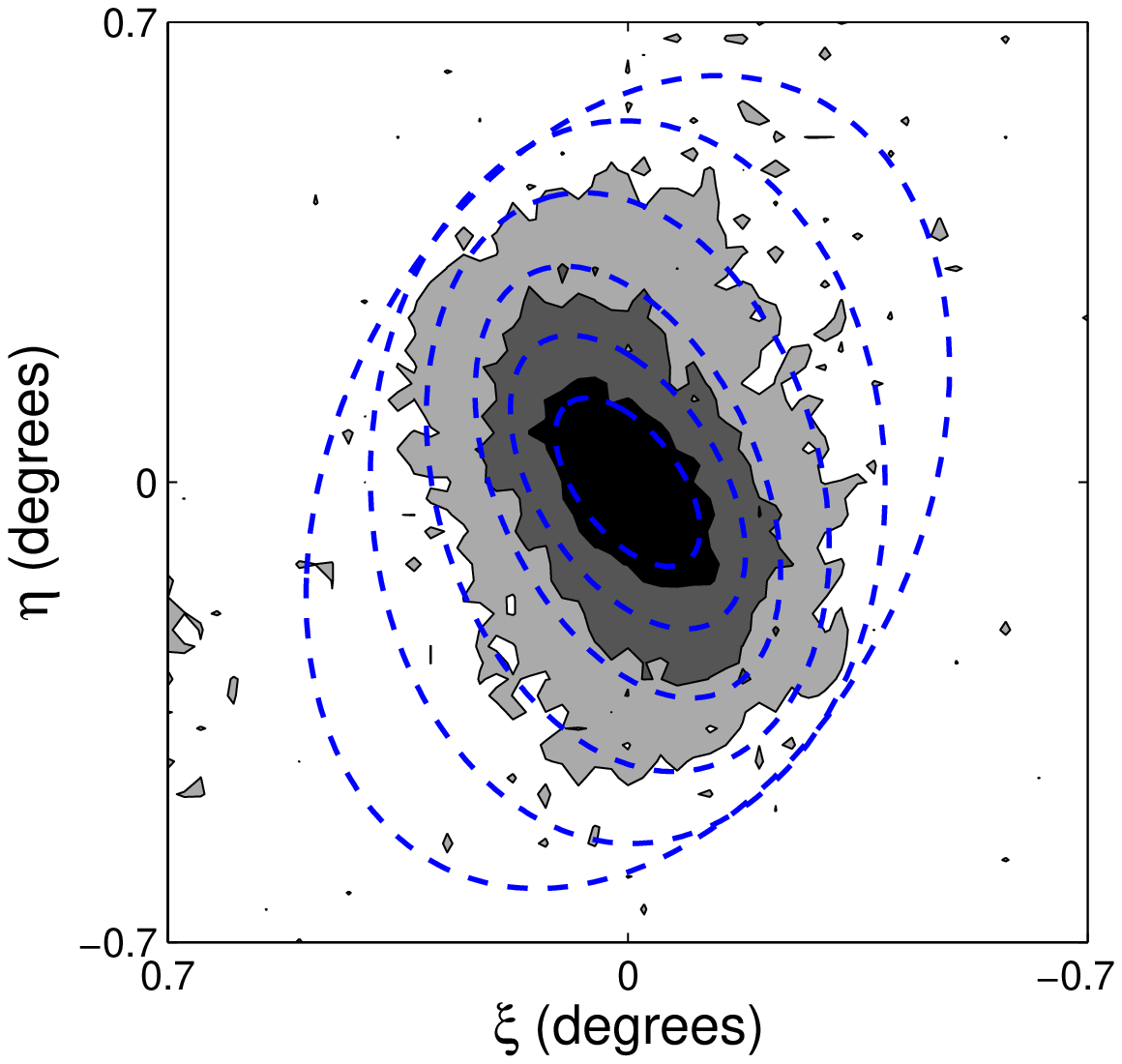}
 \includegraphics[width=8.cm]{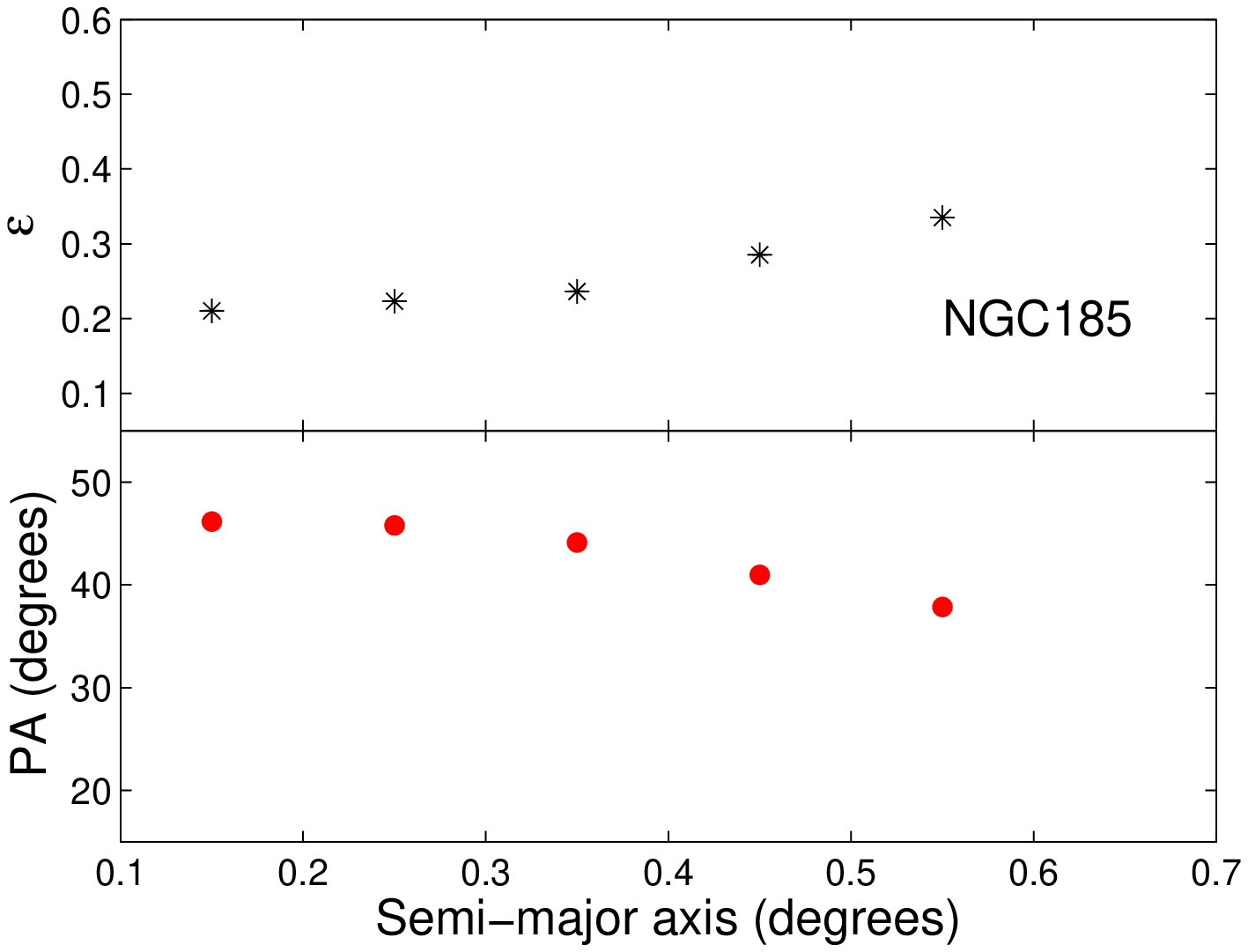}
 \includegraphics[width=8.cm]{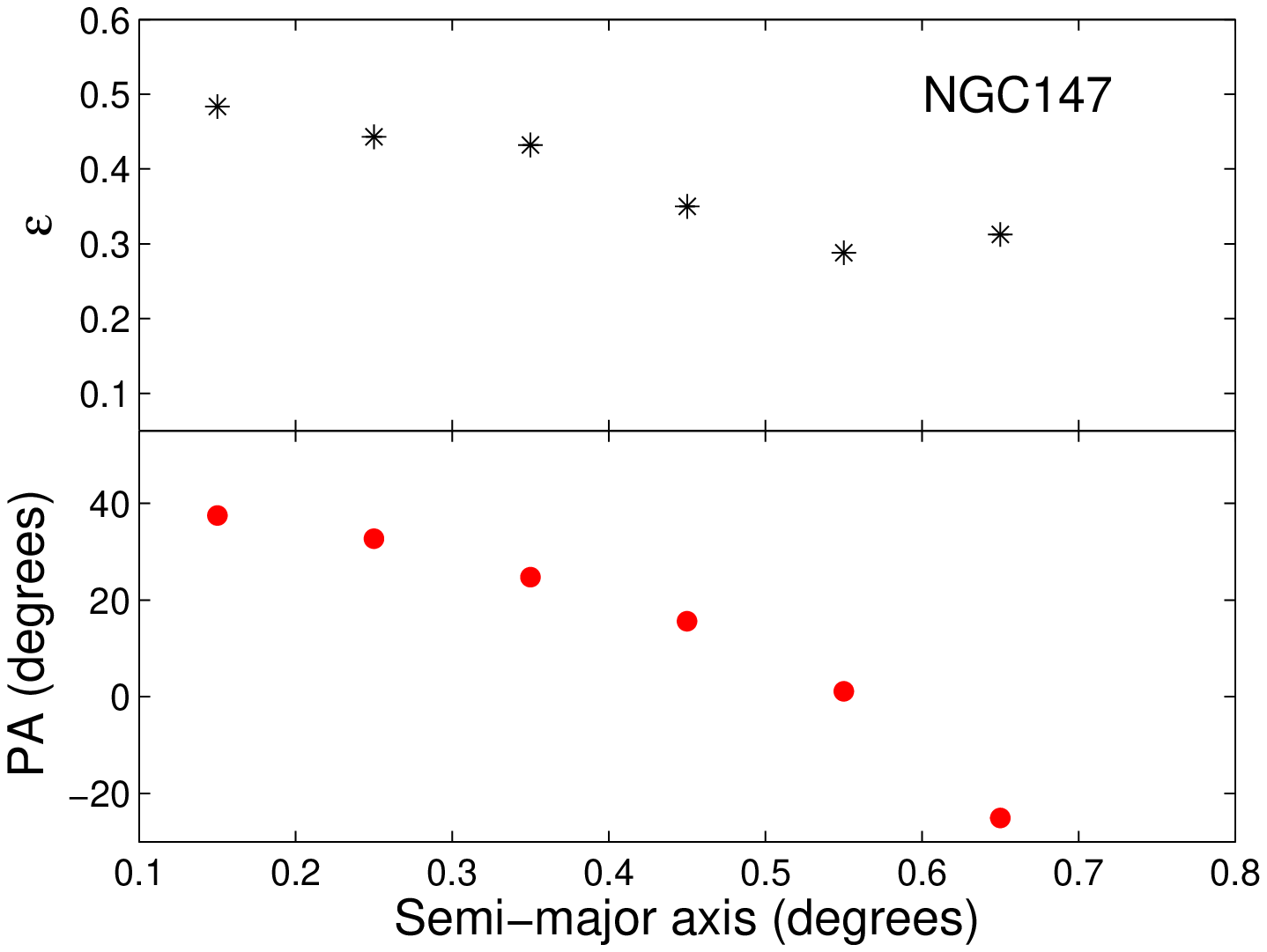}
 \caption{\emph{Top panels}: The spatial distribution of RGB stars in
   standard coordinates for NGC185 (left) and NGC147 (right). White
   stripes indicate gaps between CCDs and holes stem from saturated
   stars. We draw ellipses at 0.125, 0.25, 0.375 and 0.5~deg for
   NGC185 and at 0.2, 0.4, 0.6 and 0.8~deg for NGC147, with average
   position angle and ellipticity derived in this study, corresponding
   to the radial bins shown in Fig. \ref{cmd_spat_185} and
   \ref{cmd_spat_147}. \emph{Middle panels}: RGB density maps. The
   survey gaps/holes have been filled with artificial stars, and the
   isodensity contours are logarithmically spaced and indicate
   densities of $\sim4$, 16 and 64 stars per 1~arcmin$^2$.  Position
   angle and ellipticity are recomputed for 0.1~deg radial annuli
   (blue dashed ellipses).  \emph{Bottom panels}: Position angle and
   ellipticity as a function of distance along the semi-major axis, 
   calculated for annuli centred on each radial point.
   The errorbars are smaller than the symbols.}
\label{spatmap}
\end{figure*}

\section{Spatial analyses}     \label{analysis}
   
\begin{figure*}
  \centering
  \includegraphics[width=8.5cm]{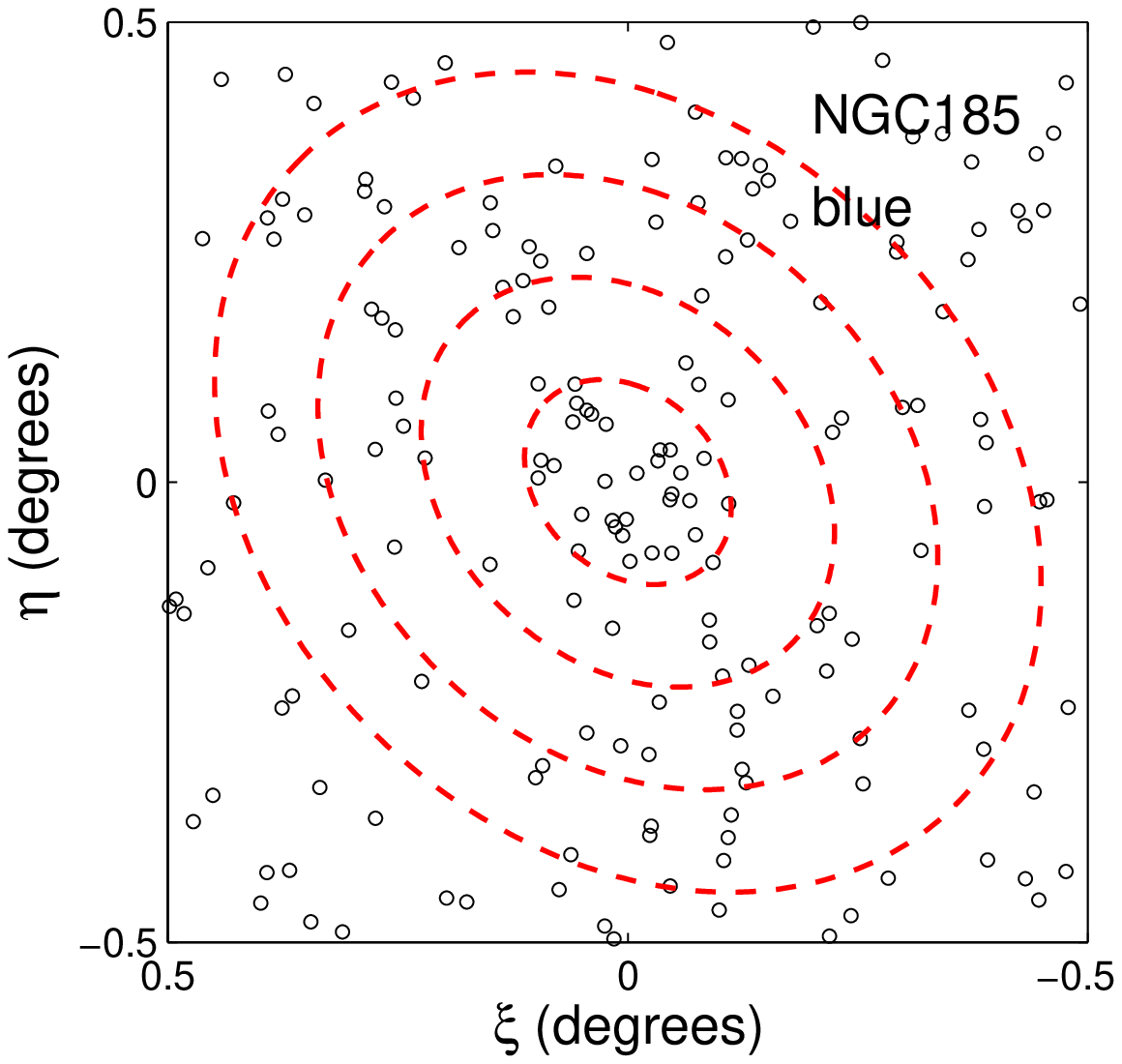}
  \includegraphics[width=8.5cm]{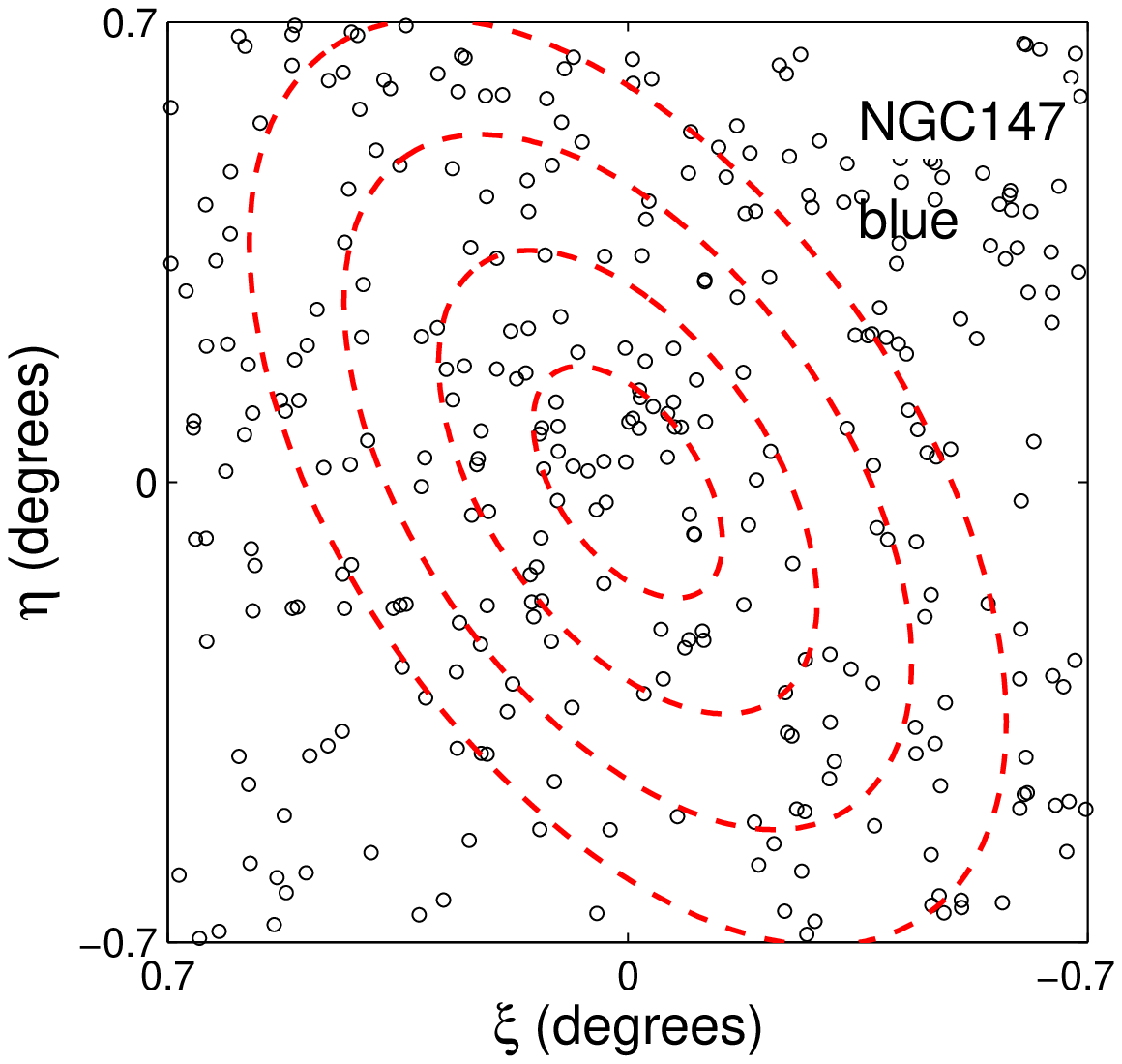}
  \caption{Spatial distribution of blue sources, consistent with being
    candidate MS stars, in each of the target galaxies.  Ellipses are
    as in the top panels of Fig. \ref{spatmap}.}
\label{spatmap2}
\end{figure*}

We analyse the spatial distribution of different stellar populations
in the observed CMDs. The top panels of Fig. \ref{spatmap} show the
distribution (in standard coordinates, centered on each dE) of point
sources falling within the RGB selection boxes drawn in Fig.
\ref{cmd_glob}.  The gaps between the CCDs in the MegaCam FoV and the
saturated stars are clearly visible.  Due to the higher central
density of NGC185, the innermost $\sim0.03$~deg is mostly unresolved.
Even in these point source density maps, one can see the significant
larger extent of NGC147 compared to NGC185 and evidence of the tidal
tails emanating in the north-west/south-east directions at large
radii.  These low surface brightness tails extend well beyond the area
analysed here and are the focus of a companion paper (Irwin et al., in
prep).

To create smoothed density maps, we artificially fill these small gaps
and holes through populating them to match source counts in adjacent
areas.  From the gap-filled distribution, we compute the RGB stellar
density per arcmin$^2$ and smooth the result with a
$2\times2$~arcmin$^2$ grid to produce the contour plots shown in the
middle panels of Fig. \ref{spatmap}.  The density contours of NGC185
are regular over its full extent, while NGC147 presents clear twisting
of the isophotes at large radius due to the tidal material. 

We use the density distribution of RGB stars to compute the position
angle $PA$ and ellipticity $\epsilon$ as a function of radius in the
two galaxies. For this, we adopt the method described in
\citet{mclaug94}, which employs the first two moments of the spatial
distribution. We begin with a circular aperture and calculate the
moments, ellipticity and position angle for the enclosed stars within
a given radius, using their formulas (8a), (8b) and (8c).  We then
iterate the process until convergence, i.e. when the shape of the
adopted aperture and that of the underlying stellar distribution are
the same.  To investigate the possibility of radial variations in the
derived quantities, we repeat this procedure for a range of radial
annuli in steps of $0.1$~deg, avoiding the innermost $\sim0.1$~deg
where crowding limits completeness.  The bottom panels of Fig.
\ref{spatmap} show the $PA$ and $\epsilon$ as a function of radius along
the semi-major axis and the resulting ellipses are shown in the middle
panels of Fig. \ref{spatmap}.  As can be seen, the $PA$ and ellipticity
vary only mildly as a function of radius for NGC185 with the
ellipticity becoming more pronounced at larger radii.  On the other
hand, as one progresses outwards in NGC147, the isophotes become
rounder and the position angle twists strongly, both a direct result
of the emergence of the tidal tails.

In the subsequent analysis, we adopt the average $PA$ and $\epsilon$
inside 0.5~deg for NGC185 and inside 0.4~deg for NGC147.  We obtain
$PA_{NGC185}=45.9^\circ \pm1.2$, $\epsilon_{NGC185}=0.22 \pm0.01$, and
$PA_{NGC147}=34.2^\circ \pm3.6$, $\epsilon_{NGC147}=0.46 \pm0.02$.
These values are in good agreement with $PA_{NGC185}=42.9^\circ$,
$\epsilon_{NGC185}=0.23$, and $PA_{NGC147}=28.4^\circ$,
$\epsilon_{NGC147}=0.44$, reported by \citet{geha10}, who derived them
from the optical photometry of \citet{kent87} spanning $\sim5'$ and
$7'$ for NGC147 and NGC185, respectively.  Ellipses corresponding to
these parameters are drawn in the top panels of Fig. \ref{spatmap}.

We also investigate the spatial distribution of sources with
$(g-i)_0<0$ which are candidate young MS populations.  Fig.
\ref{spatmap2} shows a tentative overdensity of such sources in the
central region of NGC185, while no obvious enhancement is observed in
NGC147. This is consistent with previous studies, as further discussed
in Sect. \ref{rad_sec}.  Stars falling within the two foreground
selection boxes drawn in Fig.  \ref{cmd_glob} are homogeneously
distributed, as expected for Milky Way populations. 
Sources classified as $``$extended" (and therefore rejected by our
quality cuts) are also mostly uniformly distributed but show 
enhancements in the central regions of both dEs. This
suggests that some fraction of genuine RGB stars in these parts are
misclassified as extended sources due to blending.  We will return to
this issue in the next section.

%________________________________________________________________

\begin{figure*}
  \centering
\includegraphics[width=8.5cm]{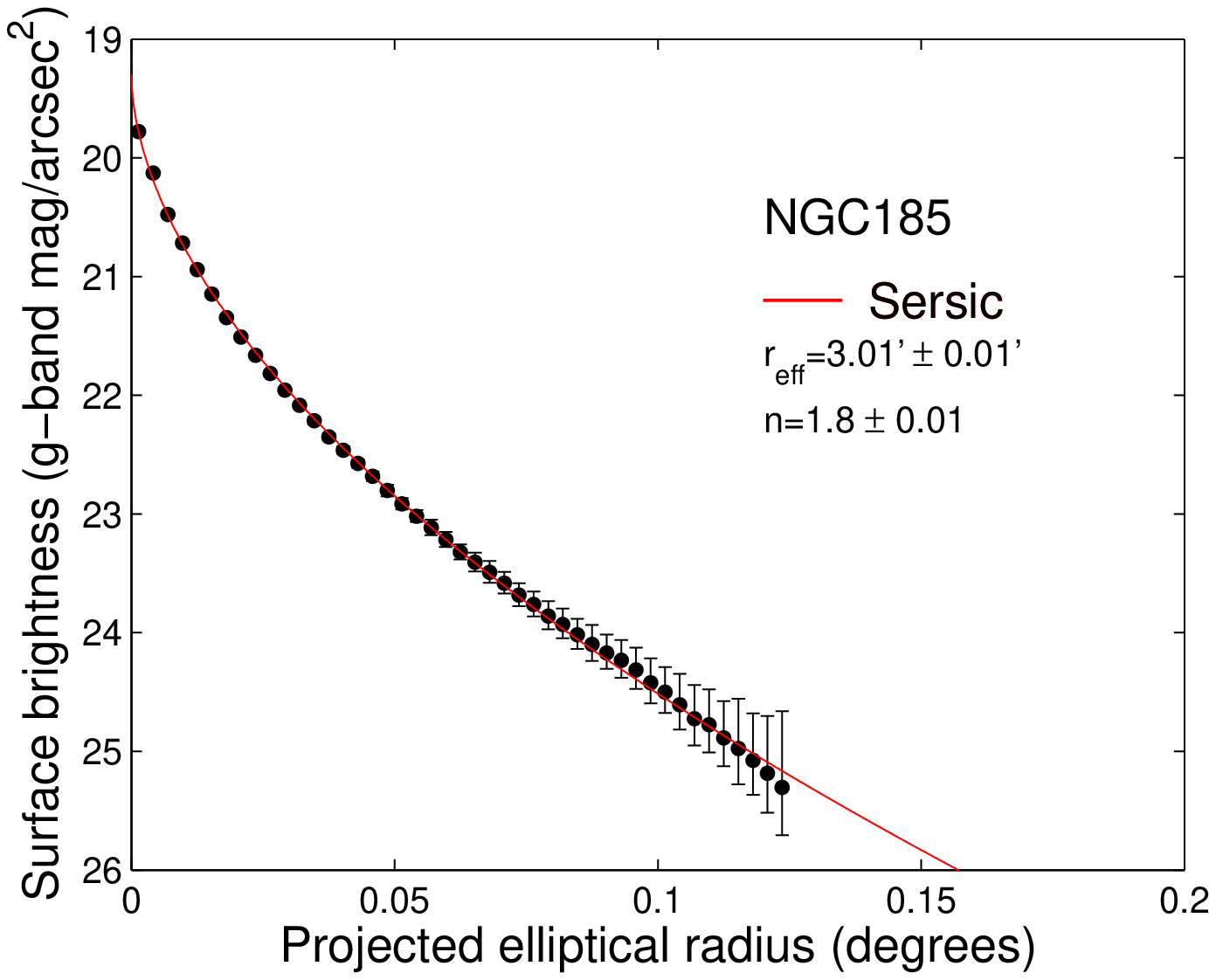}
 \includegraphics[width=8.5cm]{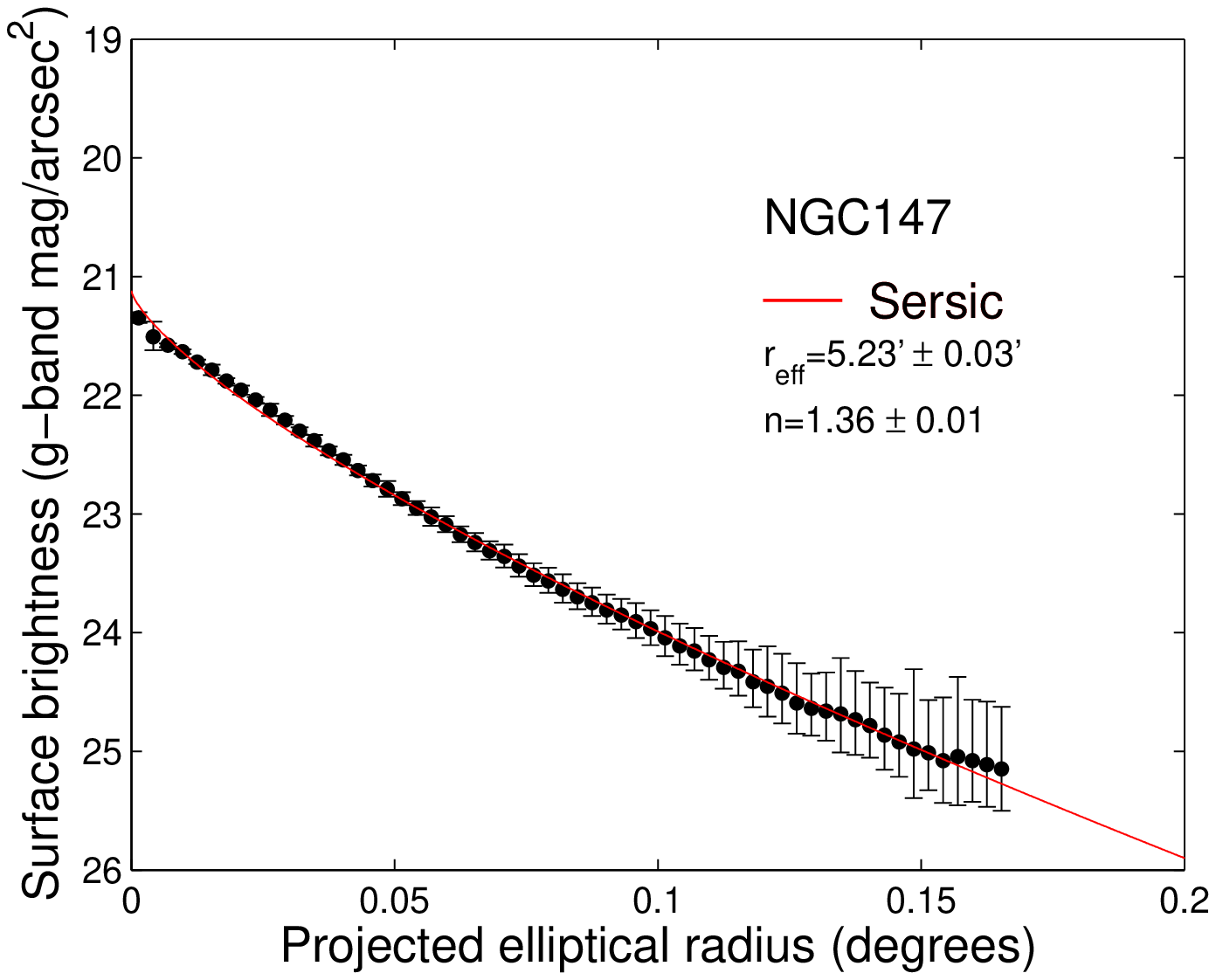}
 \caption{Surface brightness profiles ($g$-band) for NGC185 (left) and
   NGC147 (right), derived from diffuse light analyses alone. The
   best-fit Sersic profiles are overlaid, and the respective
   parameters reported.}
\label{sbprof}
\end{figure*}

\begin{figure*}
  \centering
 \includegraphics[width=8.5cm]{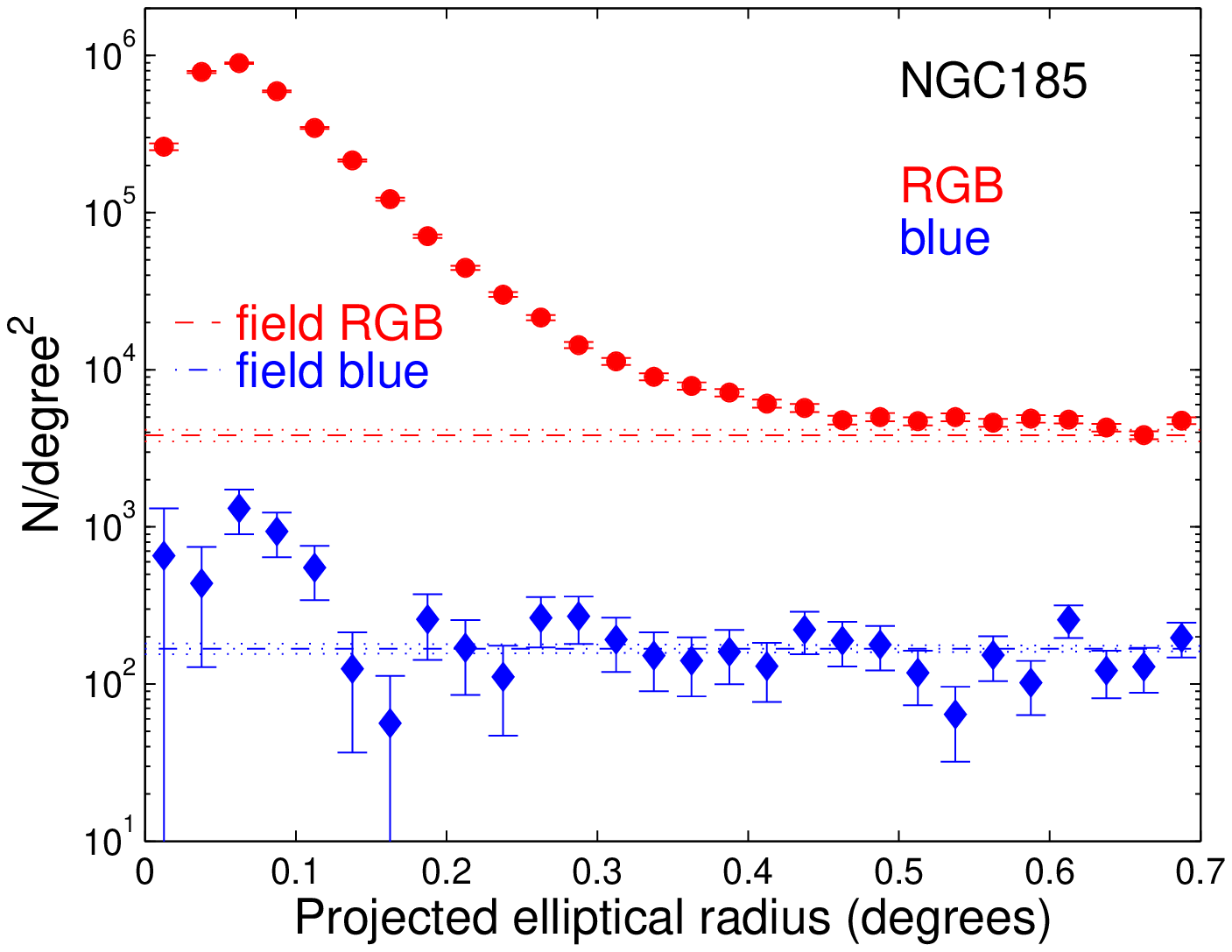}
 \includegraphics[width=8.5cm]{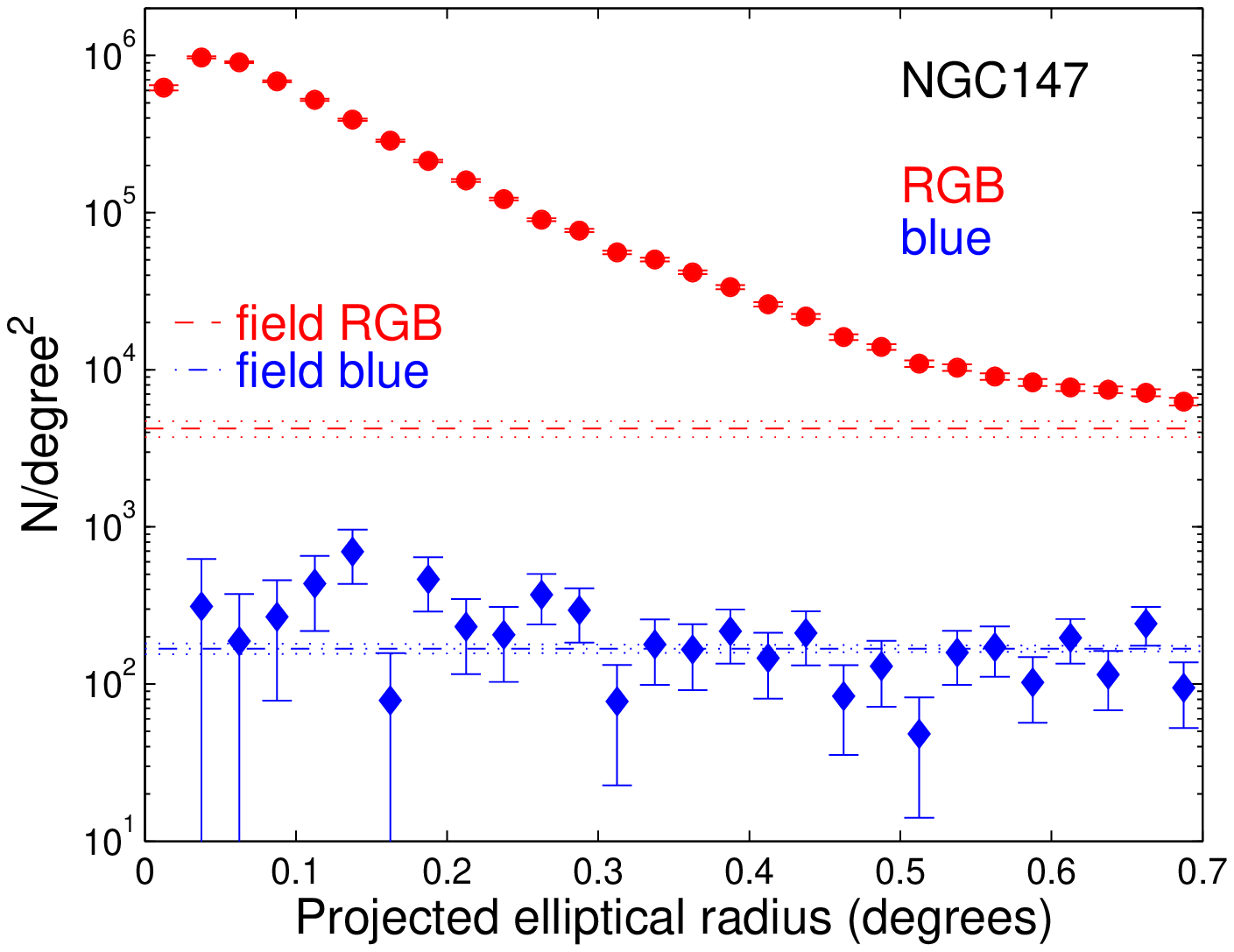}
 \caption{Radial star count profiles for NGC185 (left) and NGC147
   (right). Red circles are RGB stars, while blue diamonds are blue
   sources. The error bars are Poissonian. The red dashed line
   indicates the surface density of sources falling within the RGB box
   in the adopted field regions, i.e. the contaminant level (see Fig.
   \ref{field}), while red dotted lines indicate the dispersion around
   this value.  The blue dot-dashed line is the contaminant level for
   sources falling within the blue star selection box.}
\label{radprof}
\end{figure*}

\section{Radial profiles}   \label{prof_sec}

In this section, we derive radial surface brightness profiles for the
dEs by combining diffuse light and resolved star counts.  While
it is challenging to trace diffuse light in faint galactic 
outskirts, this is the only technique applicable to the central
regions, where high stellar crowding hampers the resolution of
individual sources (see, e.g., Fig. \ref{spatmap}).  On the other
hand, resolved star counts allow us to trace the light profiles out to
the largest galactocentric distances probed by our dataset.

\subsection{Surface brightness profiles}         \label{sb_sec}

We exploit the PAndAS images to derive a $g$-band surface brightness
profile from the diffuse light in the innermost $10'\times10'$ of each
galaxy.  After masking out bright stars, we compute the median of the
pixel values in concentric ellipses (with our adopted values of
average $PA$ and $\epsilon$).  In order to derive the sky value, we
average the mode of the pixel values in two $\sim2.5' \times 2.5'$
regions located at $15'$ on opposite sides of each galaxy, and take
the sky uncertainty to be the standard deviation of these values.  We
convert the resulting profile into mag per arcsec$^2$, and de-redden
by adopting the median $E(B-V)$ value within the considered area.

The derived surface brightness profiles are presented in Fig.
\ref{sbprof}. As pointed out before,  the central surface brightness
reached in NGC185 is higher than in NGC147 \citep[e.g.][and references
therein]{geha10}.  For NGC185, we trace the diffuse light profile out
$\sim0.12$~deg and find a best-fitting Sersic law \citep{sersic68}
with $r_{\rm eff}=3.01'\pm0.01'$ and $n=1.80 \pm 0.01$. 
The dust lane in the central $\sim0.005$~deg$^2$ of NGC185
does not significantly alter the shape of the $g$-band profile with 
respect to redder bands \citep[see also][]{kim98}. We cross-check this by measuring
the average shift between the $g$- and $i$-band profiles within 
$\sim0.005$~deg$^2$ for both NGC185 and NGC147. We use the latter
as a benchmark given the absence of dust and assume comparable ages
and metallicities, and we find the shifts to be consistent within the errorbars.
For NGC147, we trace the profile to $\sim0.17$~deg, and find 
$r_{\rm eff}=5.23'\pm0.03'$ and $n=1.36 \pm 0.01$.

\citet{geha10} used the $r$-band photometry of \citet{kent87} and the
assumption of a constant $B-V$ colour to derive $V$-band surface
brightness profiles out to radii $\sim7'$ and $5'$ in NGC185 and
NGC147, respectively.  They find slightly lower values of
$n_{NGC185}=1.76$ and $n_{NGC147}=1.04$. Their Sersic fit is performed with 
fixed values of effective radius, adopted from \citet{derijcke06}.
These values are obtained from NIR (2MASS) photometric data in the 
innermost $\sim4'$ of each dE, and are $r_{\rm eff}=1.50'$ for NGC185 and 
$r_{\rm eff}=2.04'$ for NGC147, thus at least a factor of two smaller 
than our best-fit values. 
The factor of two difference holds true when we extend the comparison 
to previous scale-length estimates resulting from fitting exponential 
functions or King profiles 
\citep{hodge63, hodge76, kent87, caldwell92, lee93b, kim98}.

\subsection{Number density profiles}         \label{rad_sec}

To construct accurate star count profiles, we need to account for the
contaminant level and its uncertainty. Fig. \ref{radprof} shows the radial surface density profiles as a
function of elliptical radius for RGB stars and blue sources in each
galaxy, with the field contamination level and uncertainty indicated. These
are calculated
as the average and standard deviation of the number densities of objects falling within the RGB and
blue selection boxes in our two adopted field regions (Fig. \ref{field}).

\begin{figure*}
  \centering
  \includegraphics[width=18cm]{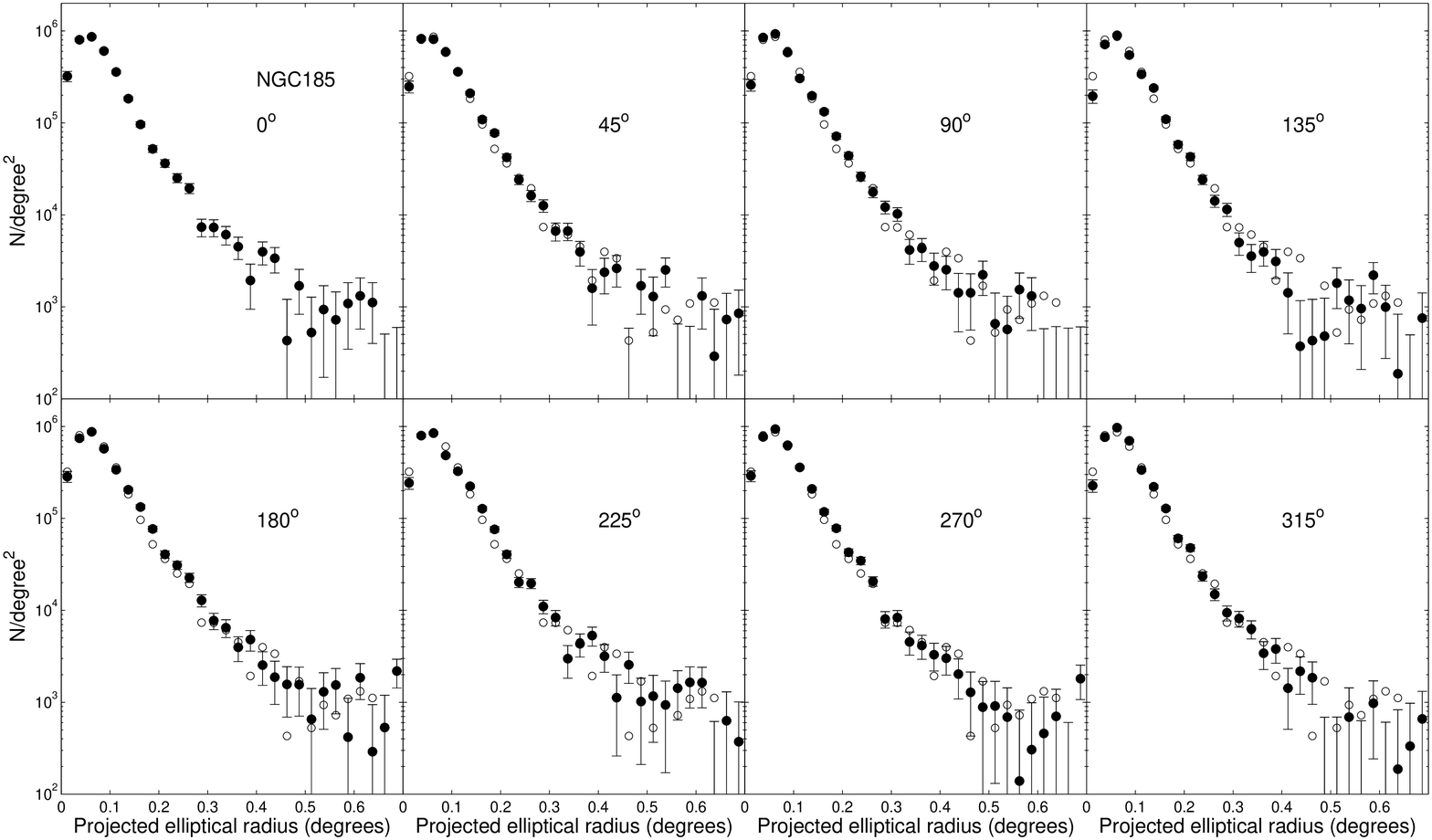}
  \includegraphics[width=18cm]{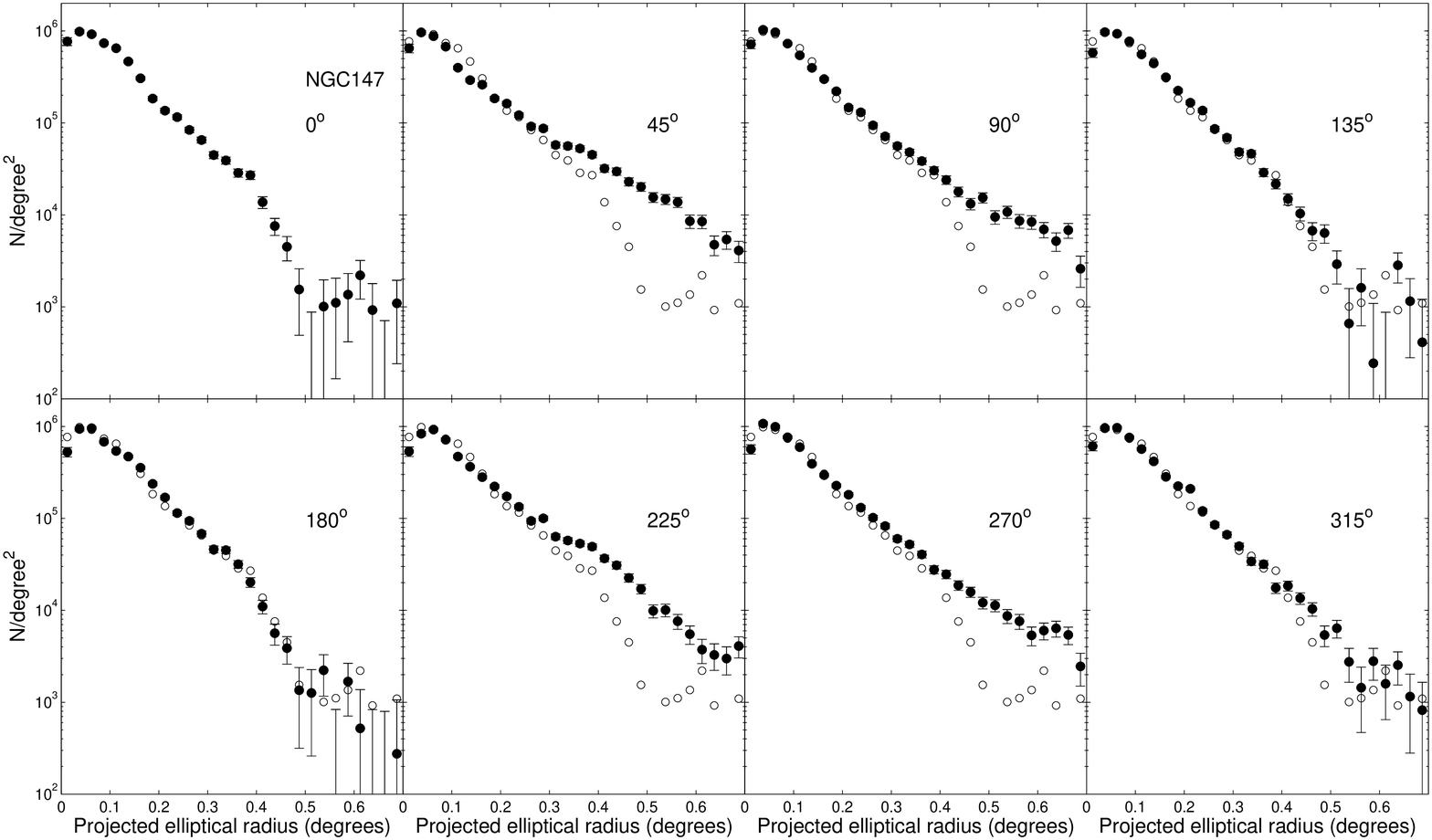}
 \caption{Radial star count profiles of RGB stars in $45^{\circ}$
    wedges as a function of azimuthal angle, for both galaxies. The
    average $PA$ and ellipticity derived in this study are assumed, and
    the profiles have been contaminant-subtracted.  The central angle
    of each wedge is measured clockwise from the north-eastern
    semi-major axis and is reported in each sub-panel. The error bars
    are Poissonian.  The $\theta=0^{\circ}$ profile is repeated in all
    other panels for reference (open symbols).}
\label{radprof_az}
\end{figure*}

Contaminants arise from Milky Way foreground stars, M31 halo
stars and unresolved background galaxies. 
While it is reasonable to expect that
unresolved galaxies and M31 halo stars are smoothly distributed across 
our FoV, the Galactic foreground may vary substantially. To check this,
we have inspected the maps of the Milky Way foreground populations 
within the PAndAS area presented by \citet[][]{martin13}. Their
Fig. 2 reveals that there are no obvious foreground overdensities in
the vicinity of NGC185 and NGC147 at magnitudes that would affect
our RGB selection box.  As
  a further check, we also use the colour-magnitude contamination model
  presented in \citet[][their formula 12]{martin13} in order to
  estimate the foreground number density in our RGB selection box, at
  the position of the target dEs.  The values we obtain from the model
  are consistent with, but lower by $\sim5-10\%$ with respect to, our
  direct estimates. This may be due to the additional sparse M31
  halo component, and/or to low-level foreground substructure.

\begin{figure*}
  \centering
\includegraphics[width=8.5cm]{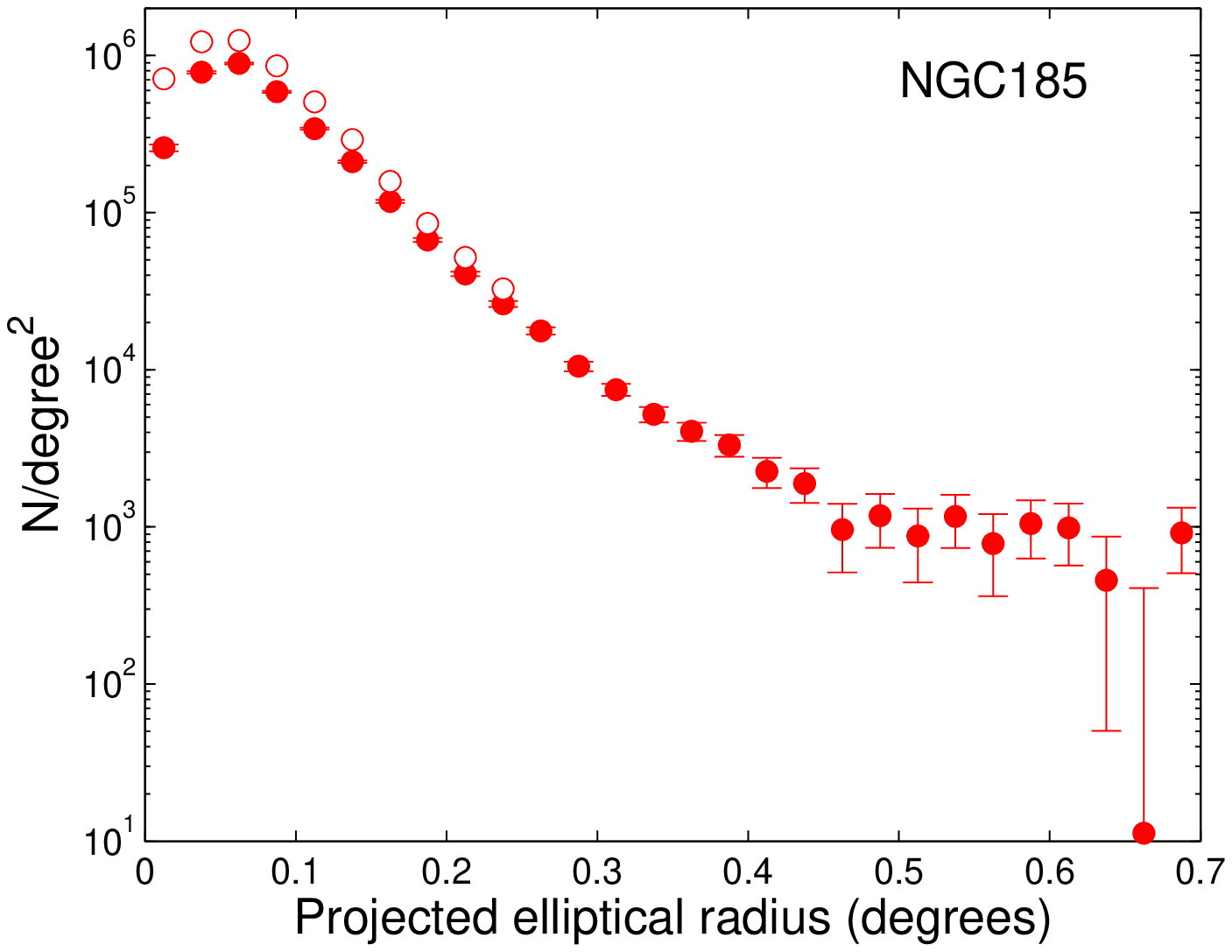}
 \includegraphics[width=8.5cm]{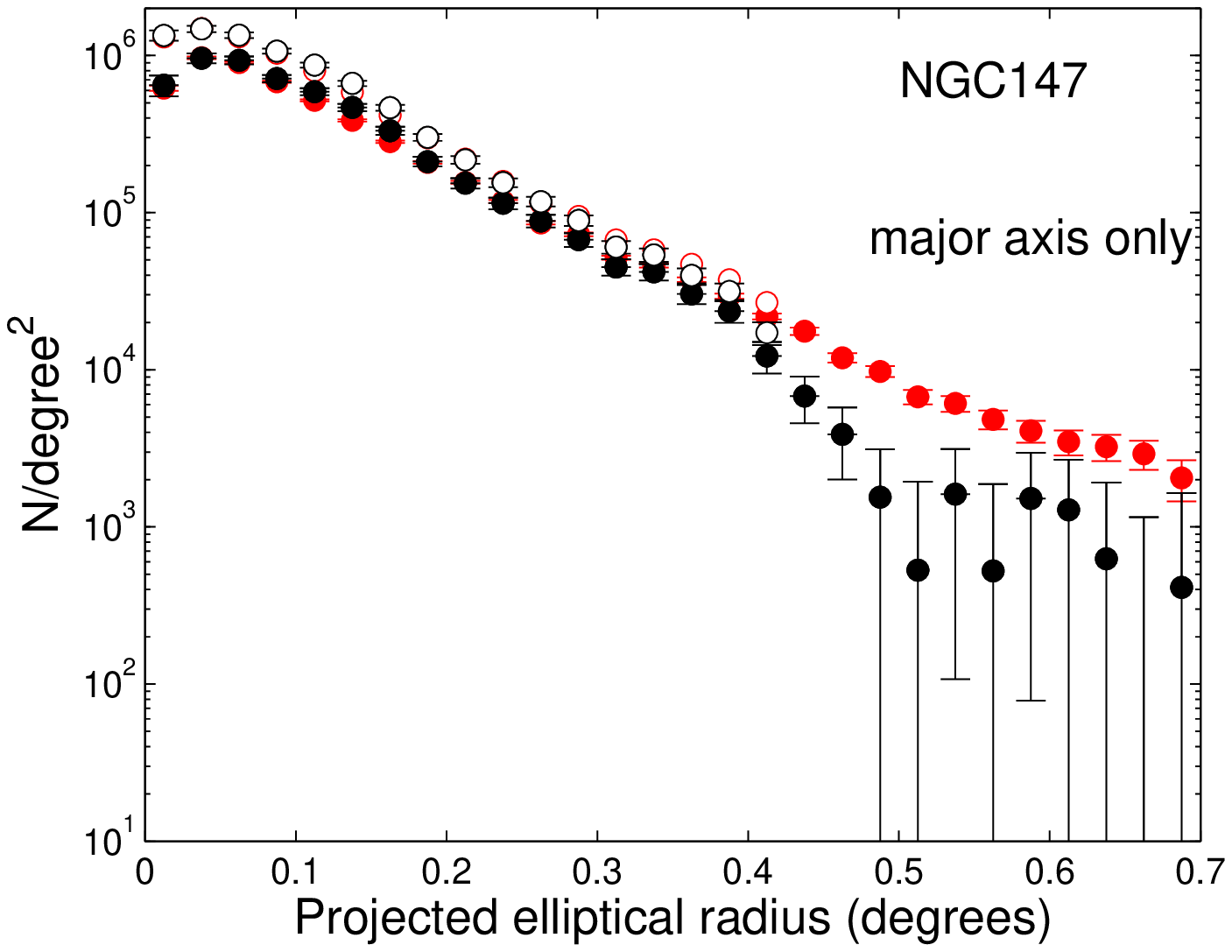}
 \caption{Contaminant-subtracted RGB star count profiles for NGC185
   (left) and NGC147 (right). Red filled circles show RGB stars only,
   while open red circles indicate the sum of both stars and extended
   sources within the RGB selection boxes.  For NGC147, black symbols
   are the same as above, but for the major axis directions only
   (i.e., the azimuthal wedges centered on $0^\circ$ and $180^\circ$
   from Fig. \ref{radprof}), to avoid the extended tidal tails.  Error
   bars are Poissonian and include the field level uncertainties.}
\label{radprof_sub}
\end{figure*}

The RGB star count profile of NGC185 approaches the contaminant level
at $\gtrsim0.5$~deg, and then plateaus at a value slightly above this.
This is likely due to low-level contamination from NGC147's tidal
tails which start to contribute in these parts.  One the other hand,
the RGB profile of NGC147 declines smoothly and remains well above the
field level out to $0.7$~deg.  Fig. \ref{radprof_az} examines how the
contaminant-subtracted RGB star count profiles vary as a function of
azimuthal angle.  Each galaxy is split into eight $45^{\circ}$-wide
wedges, with the central angle of each wedge measured clockwise from
the north-eastern semi-major axis.  For NGC185, this analysis reveals
no significant change in the radial profiles as a function of
azimuthal angle.  For NGC147, there is a clear excess of stars at
large radius in the directions of $45^\circ-90^\circ$ and
$225^\circ-270^\circ$, consistent with the presence of the tidal tails
(see Fig. \ref{spatmap}).  It can be seen that the excess stars in
these directions is what causes the azimuthally-averaged radial
density profile to remain above the field level out to $0.7$~deg.  In
directions away from the tidal tails, the RGB star count profile falls
off more steeply at large radius, and reaches the background level at
$\sim0.5$~deg.  Note that the RGB density profiles in all cases show a
drop in number counts at small radii ($\lesssim0.05$~deg), due to
incompleteness in these parts.

The radial surface density profile of blue sources is also shown in
Fig. \ref{radprof}. A mild central excess ($\gtrsim2 \sigma$) can be
seen in NGC185. While the presence of young stars is well known in
NGC185, previous observations have indicated they are mostly confined
to the inner $\sim 0.03$~deg \citep{martinez99}.  Our observations
suggest this young component may be more distributed, consistent with
inferences from some other studies \citep[e.g.][]{butler05,
  marleau10}.  While NGC147 shows a marginal overdensity of blue
sources in its central regions, this is probably not significant given
the uncertainties. In both systems, the high central densities lead to
incompleteness in the very innermost ($<0.05$~deg) regions, leading to
the turnover of the profiles.

In Fig. \ref{radprof_sub}, we show the contaminant-subtracted surface
density profile constructed using only point sources (filled symbols)
as well as point sources plus extended sources (open symbols), which
fall within the RGB selection box.  In the latter case, we only add
extended sources in when their surface density exceeds that of the
background by $>5 \sigma$, corresponding to a region inside 0.25~deg
for NGC185 and 0.4~deg for NGC147. The motivation for including these
sources is to account for stars that have been misclassified as
extended due to the severe crowding and blending at small radii.  Even
after the inclusion of these sources, the innermost regions are still
incompleteness-limited and hence we must resort to using diffuse light
in these parts.  The star count profile of NGC185 can be reliably
traced out to a
radius of $\sim0.45$~deg, corresponding to $\sim5$~kpc.  For NGC147,
we show the azimuthally-averaged star count profile as well as the
profile along the major axis only, calculated as the average of the
azimuthal profiles in centered on $0^\circ$ and $180^\circ$ (Fig.
\ref{radprof_az}).  After field level subtraction, the major axis
profile, which is virtually free from contamination by the tidal
tails, extends to $\sim0.5$~deg, or $\sim5$~kpc at NGC147's distance.
A steep break is present in the profile at
large radius, most likely due to the ongoing tidal disruption.

\subsection{Combined profiles}         \label{comb_sec}  

\begin{figure*}
  \centering
 \includegraphics[width=8.5cm]{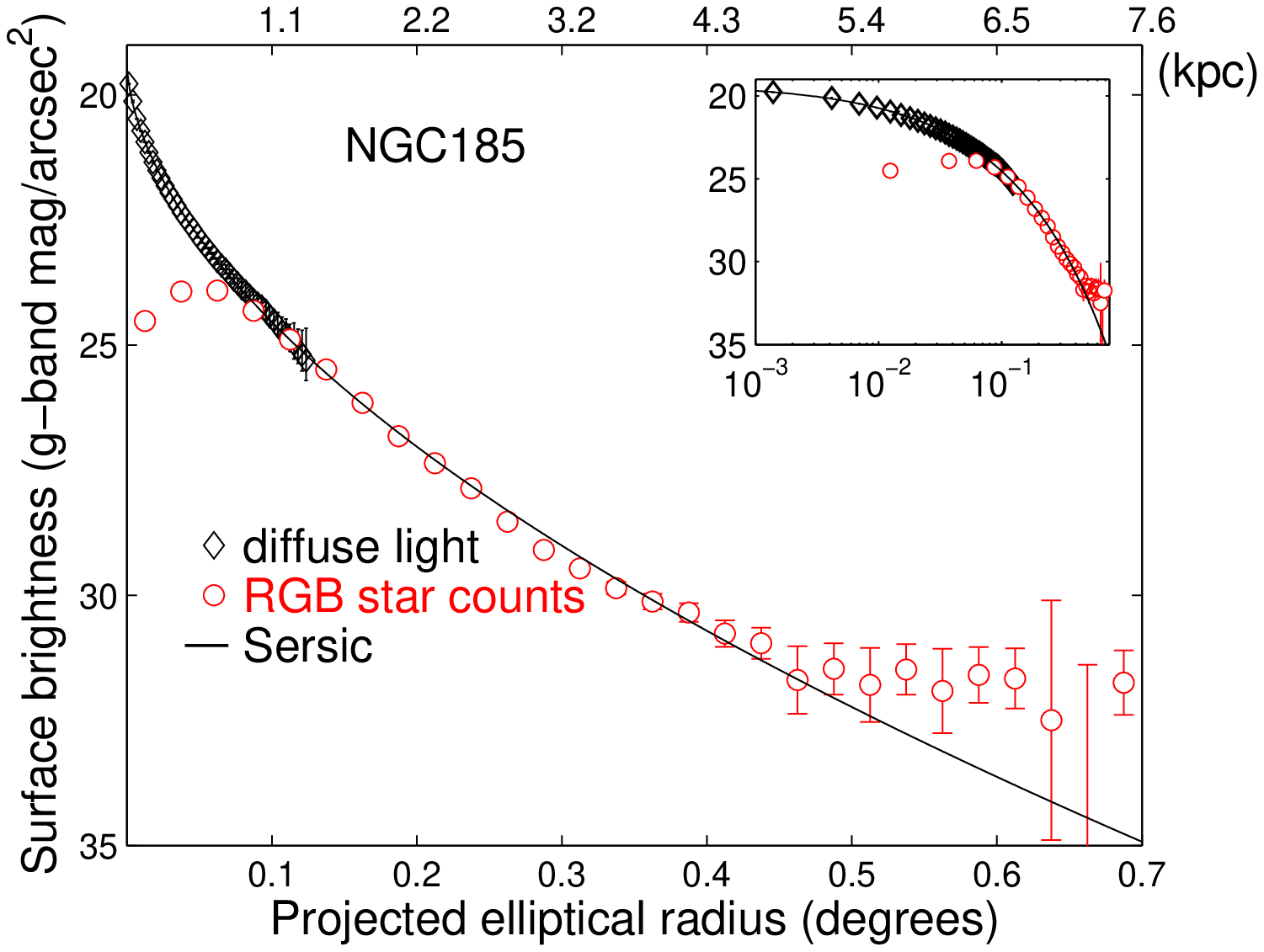}
 \includegraphics[width=8.5cm]{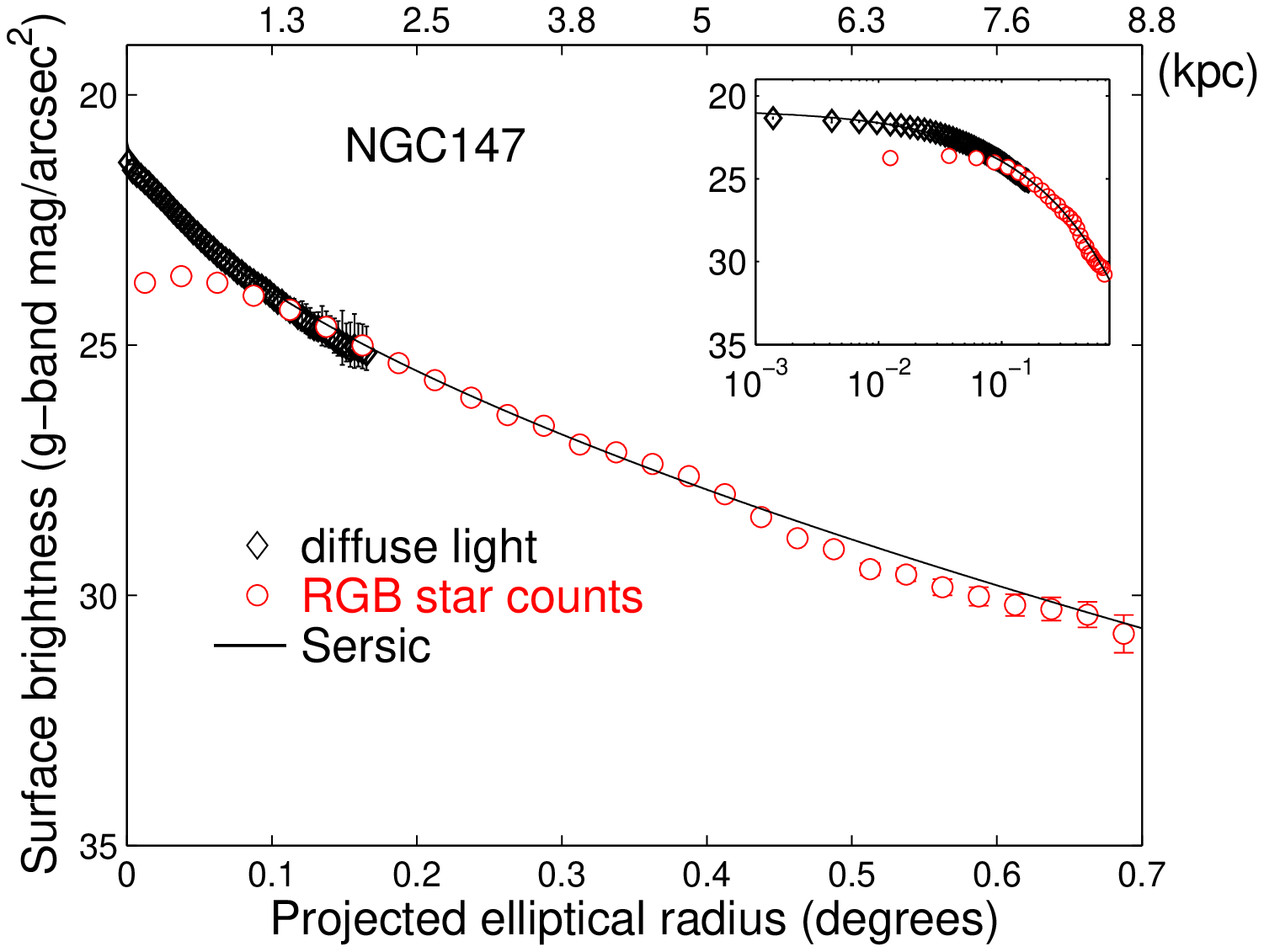}
\caption{Composite surface brightness profiles for NGC185 (left) and
   NGC147 (right) derived from merging diffuse light profiles (black
   open diamonds) with star count profiles (open red circles).  The
   best-fitting functions to the global profiles are overlaid; the
   insets show the profiles on a logarithmic scale. The upper axis
   gives the radius in physical units.}
\label{fin_prof}
\end{figure*}

We combine the diffuse light profiles shown in Fig. \ref{sbprof} with
the star count profiles $\Sigma(r)$ in order to trace the surface
brightness variation over a large radial extent. To do this, we use
the relation $\mu(r)=-2.5$~log$_{10}(\Sigma(r))+ZP$, where the
zeropoint $ZP$ is derived from the overlapping region between the
profiles (i.e., $r\sim0.1-0.13$~deg for NGC185 and $r\sim0.1-0.15$~deg
for NGC147).  We obtain $ZP=21.37$ for NGC185 and $ZP=21.36$ for
NGC147.  In constructing these profiles, we use the
azimuthally-averaged star count profiles for both dEs.  Fig.
\ref{fin_prof} demonstrates the advantage of this technique, allowing
us to trace the surface brightness from $\mu_g \sim 20-22$~mag
arcsec$^{-2}$ in the central regions to $\sim 32-33$~mag arcsec$^{-2}$ at
the largest radii.

We fit Sersic models to the composite profiles and report the results
in Tab. \ref{tab2}.  While our earlier analysis of the diffuse light
was limited to radii within $\lesssim0.2$~deg, we can fit the
composite profile out to roughly twice that radius. In particular, we
fit NGC185 in the range of $0-0.45$~deg and NGC147 in the range
$0-0.4$~deg, the latter chosen to avoid the tidal tails.  For NGC185,
the values obtained from the diffuse light and composite profiles are
consistent with each other indicating that there is no strong change
in the profile shape out to this radius. Beyond $0.45$~deg, there is
an excess of light compared to the extrapolation of the fit, but this
is likely due to contamination of these regions by NGC147's tidal
tails, as mentioned previously.  For NGC147, both $n$ and $r_{\rm eff}$ 
derived from the composite profile fit are higher than the
diffuse light-only estimates, suggesting the surface brightness
profile might flatten somewhat beyond the inner regions. Additionally,
the fit seems to provide a good description of the surface brightness
profile even beyond the radial range of the fit.  

The only other studies to probe relatively large galactocentric radii in
NGC185 and NGC147 are \citet{battinelli04a, battinelli04b}, who
analyse the distribution of giant stars in both dEs out to a radius
of $\sim0.3$~deg.  The scale-lengths $\alpha$ they obtain from exponential 
fits can be converted to effective (half-light)
radii as $r_{\rm eff}\sim1.668 \alpha$, returning $r_{\rm eff}=4.2'\pm0.1'$
for NGC185 and $r_{\rm eff}=6.8'\pm0.2'$ for NGC147. These values
can be compared to those reported in Tab. \ref{tab2}. The effective radius
for NGC185 computed by \citet{battinelli04a} is larger than ours, possibly
due to the fact that they adopt slightly larger $PA$ and $\epsilon$ 
values; the radius for NGC147 from \citet{battinelli04b} is consistent with ours.

We derive total magnitudes by integrating our Sersic composite profiles
to infinity and obtain $M_g=-15.36\pm0.04$
for NGC185 and $M_g=-16.36\pm0.04$ for NGC147. Adopting the mean RGB color
of the two dEs ($(g-i)_0\sim1.2$), we use the conversion reported in Sect.
4 of \citet{veljanoski13} to transform the CFHT/MegaCam bands into
Johnson/Cousins, and obtain $M_V\sim-15.5$ for NGC185 and 
$M_V\sim-16.5$ for NGC147. While the
previously-reported literature magnitudes for these systems are
similar (see Tab. \ref{tab1}), our revised values based on new profile
fits indicate that NGC147 is significantly more luminous than NGC185, arising from the
significant luminosity contained in its wings. 

\begin{table}
 \centering
  \caption{Best-fitting Sersic parameters to the composite profiles.}

\label{tab2}
  \begin{tabular}{lcc}
  \hline
  \hline
  & NGC185 & NGC147 \\
\hline

$\mu_{g,\rm eff}$ (mag) & $22.81\pm0.03$ & $24.15\pm0.03$ \\
$r_{\rm eff}$ (') & $2.94\pm0.04$ & $6.70\pm0.09$ \\ 
$r_{\rm eff}$ (kpc) & $0.53\pm0.01$ & $1.41\pm0.02$ \\ 
$n$ & $1.78\pm0.02$ & $1.69\pm0.03$ \\
$\chi^2_{\rm red}$ & 0.16 & 0.13 \\
$M_V$ (mag)& $-15.5\pm0.04$ & $-16.5\pm0.04$ \\

 \hline
 \hline
\end{tabular}
\end{table}

%________________________________________________________________

\section{Metallicity distribution functions} \label{mdfss}

The mean locus and width of the RGB are widely used as indicators of
the metal content of a galaxy, given that they are more sensitive to
metallicity than to age \citep[e.g.][]{vandenberg06}.  If it is
assumed that RGB stars are coeval, the width of the RGB can be
interpreted as the spread in metallicity within a galaxy.  We adopt
the updated 2012 version of the Dartmouth stellar isochrones
\citep{dotter08, mcconnachie10} with a fixed age of 12~Gyr and
[$\alpha$/Fe] $=+0.0$ to derive the photometric MDFs for each galaxy.
We only consider RGB stars bounded by the [Fe/H]$= -2.5$ and $-0.3$
isochrones, with magnitudes $i_0<23$ and fainter than the TRGB.
Metallicities are computed by linearly interpolating between
isochrones with a [Fe/H] spacing of 0.2~dex.

\begin{figure*}
  \centering
 \includegraphics[width=8.5cm]{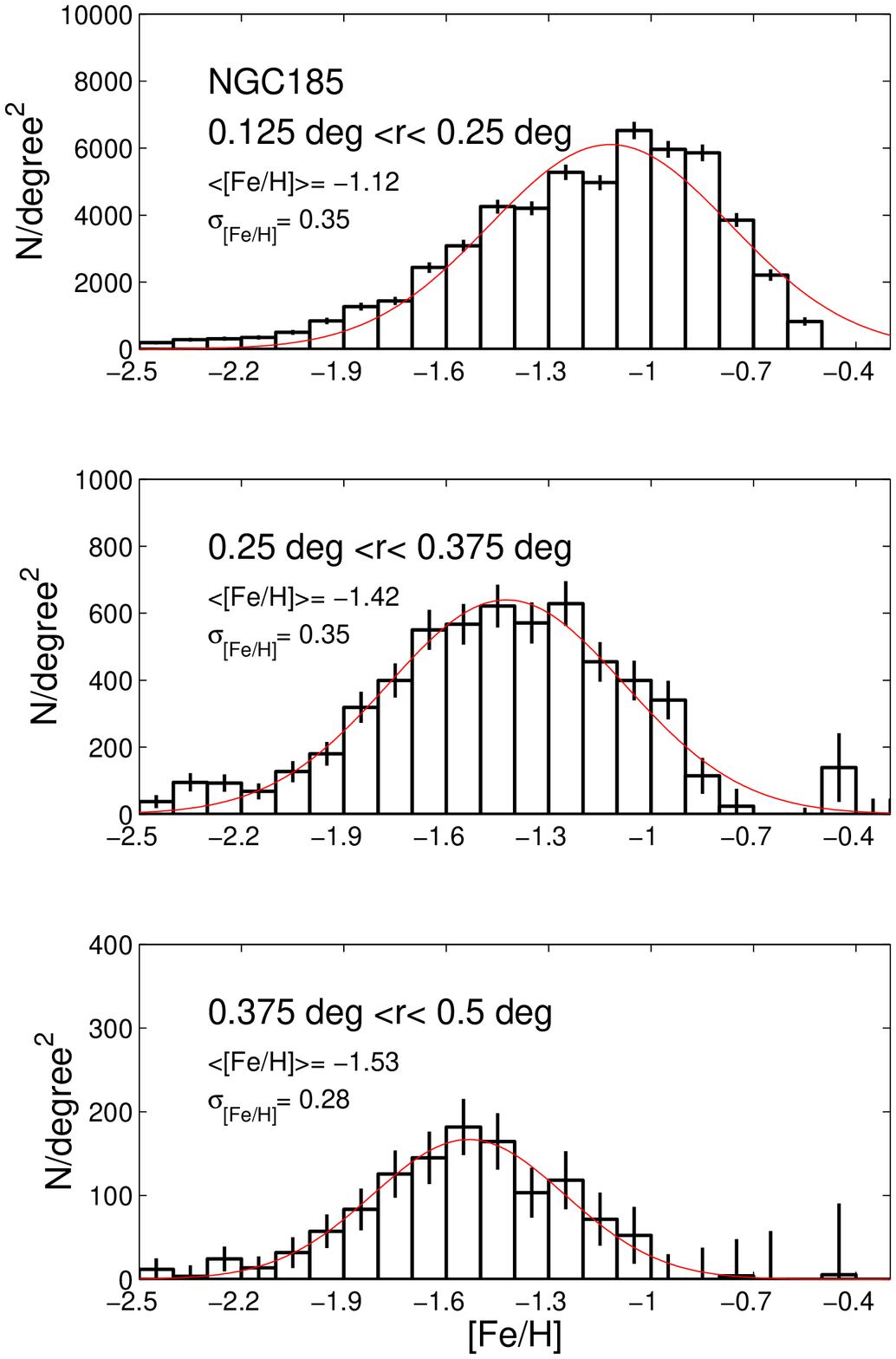}
 \includegraphics[width=8.5cm]{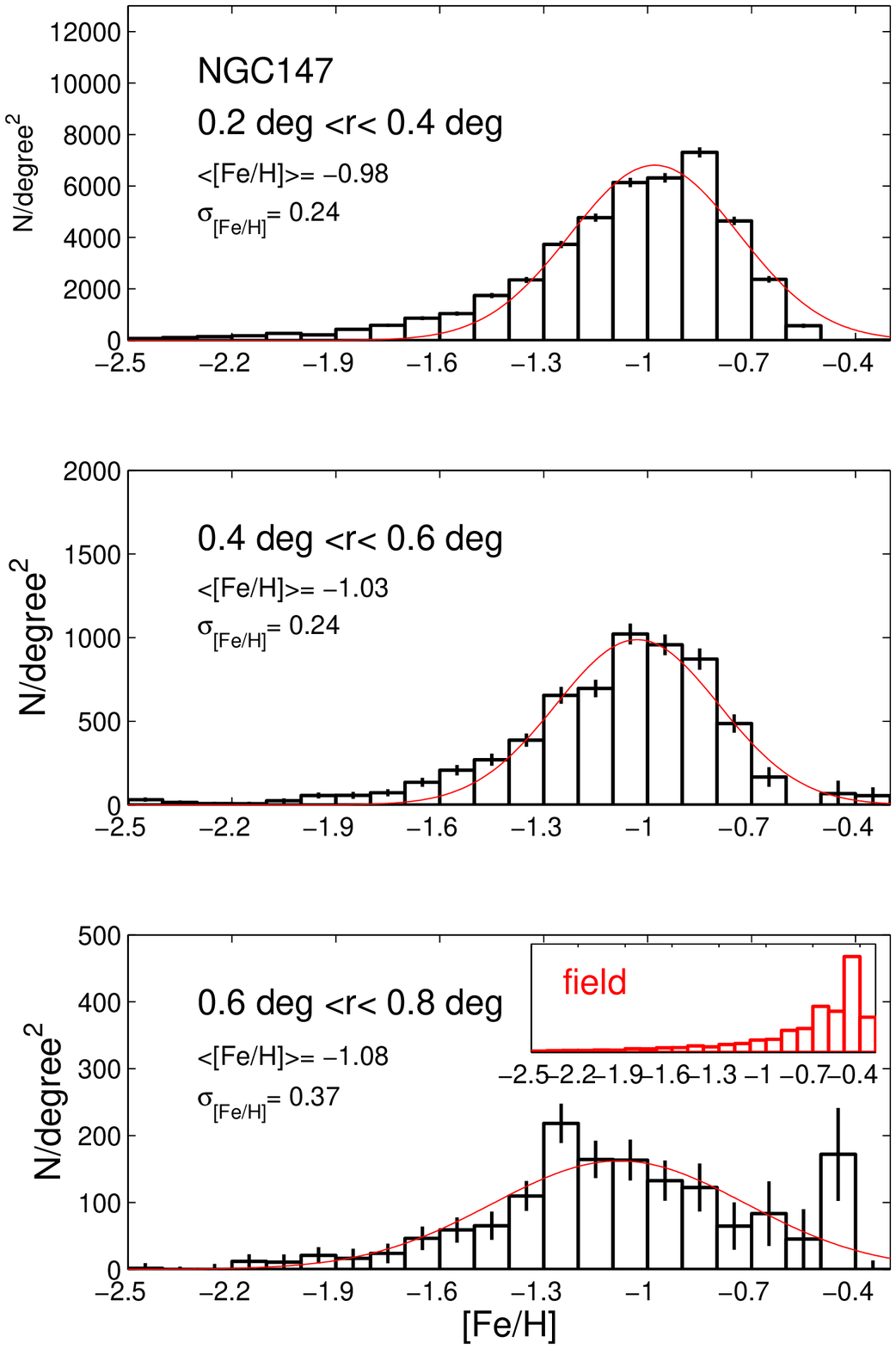}
 \caption{Photometric MDFs (per unit area) as a function of radius for
   NGC 185 (left) and NGC 147 (right), derived via isochrone
   interpolation assuming a fixed age (12 Gyr). We only consider RGB
   stars with $i_{0,TRGB}<i_0<23$.  The field contamination has been
   subtracted (see text for details). The radial bin, mean metallicity
   and metallicity dispersion (obtained from the best-fitting
   Gaussian, red curve) are reported for each sub-panel. The errorbars
   are Poissonian.  }
\label{spat_mdfs_sub}
\end{figure*}

The influence of contaminants again needs to be taken into account when
interpreting the MDFs.  To assess this, we construct MDFs for point
sources in the adopted field regions that fulfill the same
colour-magnitude selection criteria as used for RGB stars in the dwarf
galaxies. As before, it is reasonable to assume that these sources are
not genuine RGB stars but foreground/background contaminants.  We
subtract the contaminant MDFs (after area scaling) from the MDFs of
the target galaxies and fit a Gaussian to quantify the mean
metallicity and dispersion.  For NGC185, we find $<$[Fe/H]$> = -1.08$
and $\sigma_{[Fe/H]}= 0.39$ (within 0.5~deg), 
while for NGC147 we find $<$[Fe/H]$> =-0.96$ and $\sigma_{[Fe/H]}= 0.30$
(within 0.8~deg).  These global estimates may be underestimates
of the true values since the RGB sample is incomplete at small radii, where
the most metal-rich stars are likely to lie.

The MDFs as a function of radius are presented in Fig.
\ref{spat_mdfs_sub}, using the same radial bins as in Fig.
\ref{cmd_spat_185} and \ref{cmd_spat_147}. We do not consider the
innermost bin due to incompleteness.  In the same figure, we report
the mean metallicity and metallicity dispersion obtained from the
best-fitting Gaussian (red lines) in each bin.  The uncertainty in the
individual RGB metallicity values is much smaller than the derived
metallicity dispersion.  We do not include the rejected extended
sources in computing the MDFs as they comprise only a small fraction
of the total source counts in the radial bins considered.

The two dEs show very different radial trends in their MDFs. The mean
metallicity in NGC 185 decreases significantly as a function of radius
and presents a gradient of $\Delta{{<[Fe/H]>} \over R} \sim
-0.15$~dex/kpc over radii spanning 0.125 to 0.5~deg. In terms of
physical radii, this corresponds to $\sim 1.3-5.4$~kpc, or
$\sim4-9~r_{\rm eff}$.  The global metallicity for NGC 185 found above
($<$[Fe/H]$> = -1.08$) is higher than in any of the panels shown in
Fig. \ref{spat_mdfs_sub}, indicating that this gradient extends to
stars inside 0.125~deg.  The width of the MDF becomes slightly
narrower beyond 0.375~deg, varying from $\sigma_{\rm [Fe/H]} = 0.35$
to 0.28.  On the other hand, the mean metallicity varies only slight
across NGC147, amounting to a gradient of $\Delta{{<[Fe/H]>} \over R}
\sim -0.02$~dex/kpc from 0.2 to 0.6~deg.  This corresponds to $\sim
2.5-6.3$~kpc, or $\sim2.5-4.5$~$r_{\rm eff}$.  In the outermost bin,
stars from the tidal tails contribute (see Fig. \ref{radprof_sub}) but
their stellar density is low and comparable to the contaminant count.
The peak in the MDF at $\sim -0.4$ dex probably arises from residual
contaminants as it corresponds to a similar peak seen in the field MDF
(see inset).  If we make the extreme assumption that all stars with
[Fe/H]$>-0.7$ in our outermost bin are contaminants, the best-fitting
mean metallicity and metallicity spread change to $<$[Fe/H]$> =
-1.13\pm0.28$.

Photometric metallicity estimates can suffer from
uncertainties due to the underlying assumptions (i.e., a coeval old
population) and due to the choice of theoretical libraries. While
this limits their ability to 
return precise \emph{absolute} values, the derivation of
\emph{relative} values, e.g. radial trends, should be robust.  We
consider the possible biases stemming from the neglect of
intermediate-age populations, which are known to be present in the
target dEs \citep[e.g.][see also our Fig. \ref{cmd_glob}]{nowo03,
  battinelli04a, battinelli04b, butler05, davidge05, kang05}.  Both
intermediate-age RGB and old first-ascent AGB stars will populate the
blue side of the RGB ([Fe/H] $\lesssim -1.0$~dex) and thus will
artificially increase the metal-poor fraction in the observed MDFs.
If these populations are more centrally concentrated than
the bulk of the old populations in these galaxies, as may be the case
(\citealt{battinelli04a, battinelli04b}), they will also skew
photometric metallicity gradients.  Their number counts thus mainly
affect the MDFs at small radii ($<0.125$~deg for NGC185 and $<0.4$~deg
for NGC147).  If we assume, for example, that $20\%$ ($50\%$) of the
stars blueward of [Fe/H] $\sim -1.0$~dex are intermediate-age RGB
stars, we obtain best-fitting values for the innermost radial bin of
$<$[Fe/H]$> = -1.08$ ($-0.96$) for NGC185 and $<$[Fe/H]$> = -0.95$
($-0.89$) for NGC147. These values are more metal-rich than the
original values by $\sim0.04$~dex ($\sim0.16$~dex) for NGC185, and
$\sim0.03$~dex ($\sim0.09$~dex) for NGC147, thus not significantly
affecting our conclusions about the difference in metallicity gradient
between the two galaxies.

%________________________________________________________________

\section{Discussion} \label{disc}

Our wide FoV observations have allowed us
to survey the outer regions of NGC185 and NGC147, well beyond previous
studies, and establish that the stellar extents of these
systems are greater than the literature estimates. Until our work, the widest 
FoV CCD observations of these dEs
have been performed by \citet{battinelli04a, battinelli04b},
covering $42'\times28'$ centered on each galaxy with a limiting 
magnitude of $\sim2$~mag below the TRGB in the $I$ band.
The spectroscopic samples of \citet{geha10} and \citet{ho14} reach
$\sim15'$ on each side of the major axis.
Our analysis maps the brightest $\sim3$~mag
of the RGB over a region of $\gtrsim1$~deg$^2$ around the target galaxies.
After correcting for contaminants, we have shown that 
NGC185's population of RGB stars  can be mapped out to 
at least 0.5~deg (or $\sim5$~kpc). In directions which avoid the tidal tails (i.e., along 
the major axis), NGC147 can be mapped out to a similar radius. The results
of fitting Sersic models to both the diffuse light-only profiles and
the composite profiles yield effective radii significantly larger
than most previous literature estimates by a factor of $\sim2$,
and reveal NGC147 has significant luminosity in its extended
wings. We argue that the larger effective radius and shallower
surface brightness profile of NGC147 with respect to NGC185 are
a result of its strong tidal interaction, and at this stage
it is not possible to trace its originally undisturbed profile.

We find that stars in the outskirts of NGC 147 are 
distributed in a highly asymmetric fashion,
with prominent tidal tails elongating in the north/north-west to
south/south-east direction. This is in stark contrast to the regular
isophotes which characterise the outer regions of NGC185.
We cannot rule out the possibility
that NGC185 has tidal tails which lie below the surface brightness sensitivity
of the PAndAS survey. Tidal effects only become 
evident at radii where the local crossing time is longer than the 
time elapsed since a tidal interaction \citep{penarrubia09},
and thus signs of the interaction are erased quicker in the inner
regions of a galaxy. The tails in NGC147 become visible at a radius of 
$\sim5$~kpc and at a surface brightness of $\sim27.5$~mag arcsec$^{-2}$.
The velocity dispersion of NGC147 is $\sim1.5$ higher than that 
of NGC185 \citep{geha10}, thus we would expect to detect possible
tails in NGC185 at $\sim7.5$~kpc. Within this galactocentric distance 
we do not observe azimuthal variations in its radial profile (Fig. 
\ref{radprof_az}), and if tails were present beyond this radius 
they would be consistent with the field level at a surface brightness 
of $\sim32$~mag arcsec$^{-2}$.
We conclude that, if a tidal distortion is present in NGC185, it
would be significantly weaker than what observed in NGC147.

These different outer structures may hold clues about the interaction
histories of the galaxies.  Some recent work has suggested 
that the two dEs do not form a bound pair based on their phase-space distribution
functions together with the timing argument \citep{watkins13} 
or on gravitational arguments \citep{evslin13}.  Comparing to
cosmological simulations, \citet{teyssier12}
argue that for NGC185 a past passage through the Milky Way is probable.
On the other hand, \citet{fattahi13} suggest they could be a physical pair
based on their proximity, similar luminosities and radial velocities.  
With the updated values of radial velocities and total mass from 
\citet{geha10} and the distances from \citet{conn12}, we find the ratio
between potential energy and kinetic energy in this system to be $\sim1$,  
thus the extant data cannot give a conclusive result on whether the two dEs are bound
to each other.   Any model for the orbital evolution of the two galaxies
needs to explain why prominent tidal tails are visible around
NGC147 and not NGC185.   A possible scenario is one where
the two dwarfs were bound to each other in the past and, at their
last pericentric passage around M31,  NGC147 was closer to M31 than NGC185 was.
A separation between the dEs of a few tens of kpc could result in 
only NGC147 experiencing a strong tidal interaction with M31,
as demonstrated by the simulations of Arias et al., submitted.

\citet{geha10} derived spectroscopic values for individual
RGB stars along the major axis out to 0.25~deg for both galaxies,
and more recently \citet{ho14} re-analysed the same sample 
with updated spectroscopic calibrations 
to derive MDFs and radial metallicity gradients.
Our global metallicity and metallicity dispersion for NGC185, dominated  
by stars in the innermost $\sim0.2$~deg,
are only slightly lower than the values reported by \citet{ho14}
and can likely be explained by incompleteness in our star count data
at small radii. This is 
reassuring when considering the uncertainties involved in our
photometric method.
On the other hand, \citet{ho14} find that the mean metallicity for NGC147
is about twice the value derived by \citet{geha10}, while our value agrees with 
the latter. \citet{ho14}  argue that the higher value found in their more
recent analysis would be consistent with the fact that NGC147 has
experienced considerable prolonged star formation.  Considering
metallicity gradients, 
\citet{battinelli04b} do not find a gradient for NGC185 from the
ratio of C- to M-stars, while \citet{battinelli04a} find that for NGC147 [Fe/H] decreases 
by $\sim0.4$~dex within the innermost $\sim0.25$~deg.  On the
other hand, 
\citet{ho14} find a strong metallicity gradient for NGC185 
and none for NGC147.  This latter result is in line with our results and we
have been able to show that these trends continue to larger radii than
previously considered.

Metallicity gradients have often been used in the past
to constrain the origin of dE galaxies \citep[e.g.][]{koleva09, koleva11, spolaor09}.
The formation of metallicity gradients in isolated dwarf galaxies is generally
understood to be a physical process driven by internal feedback
and by the stellar orbital anisotropy 
\citep{marcolini08, valcke08, revaz09, stinson09, pasetto10}, 
but their longevity can critically depend on the galaxy's evolutionary history.
The rotating dwarfs simulated in \citet{schroyen11} display
shallow gradients, while \citet{schroyen13} find that
non-rotating dwarf galaxies can easily retain their gradients over many Gyr.
Ram-pressure stripping and tidal interactions may also
play a role in the evolution of dwarf galaxies, especially in transforming a dIrr into a dE
\citep[e.g.][]{moore98, mayer06}, however models to date have not explored the
consequences of these processes for metallicity gradients.  
\citet{geha10}'s analysis shows NGC147 and NCG185 continually rising
rotation curves out to at least 0.25~deg.
The ratio between the maximum rotation velocity and the velocity dispersion 
$V_{max}/\sigma$ for the two dEs is similar ($\gtrsim0.9$ for NGC147 and $\gtrsim0.65$ for NGC185), 
and yet we find that they have completely different metallicity gradients. \citet{leaman13}
compared dE and dIrr dwarfs within the LG, and suggested a correlation between
$V_{max}/\sigma$ and metallicity gradient strength, with lower $V_{max}/\sigma$ corresponding
to steeper gradients. Although this trend agrees with the simulations of \citet{schroyen11} 
(but see \citealt{ho14}), it
is at odds with the properties of the dwarfs studied here.  The likely reason for this is
the strong tidal disruption that NGC147 is experiencing and the rearrangement of
material resulting from this. 

%________________________________________________________________

\section{Conclusions} \label{concl}

We have performed a deep wide-field analysis of the resolved
stellar populations in the dE companions of M31,  NGC185 and NGC147.
Using data from the PAndAS survey, we have constructed
 CMDs which reach down to $\sim3$~mag below the TRGB and allow
 us to conduct the first detailed exploration of the structure and stellar content of the peripheral
 regions of these systems.

Our results can be summarized as follows:

\begin{itemize} 

\item RGB star count maps of NGC185 exhibit regular, elliptical isophotes out to the
furthermost radii probed, while NGC147 shows prominent isophotal twisting and 
the emergence of extended
tidal tails (see also Irwin et al., in prep.);

\item we derive composite surface brightness profiles by combining diffuse light
in the central regions of the dEs with resolved stellar counts in the outskirts, enabling
us to trace them to
magnitudes as faint as $\mu_g \sim 32$~mag arcsec$^{-2}$. The profiles of both galaxies
can be fit by single Sersic profiles: NGC185 shows a much steeper profile with a higher central
surface brightness with respect to NGC147. Surprisingly, the effective radii we 
derive are a factor of $\sim2$ larger than most previous literature estimates based on
smaller FoVs. We use our new profiles to recalculate the integrated magnitudes of the two dEs, 
concluding both are more luminous than previously recognised and 
that NGC147 is roughly a magnitude brighter NGC185;

\item we derive photometric metallicity distribution functions for the RGB stars. 
An analysis of the radial variation in [Fe/H] reveals a pronounced negative gradient in NGC185
($\sim-0.15$~dex/kpc in the range $0.125<r<0.5$~deg, or over $2.7$~kpc) as opposed 
to NGC147, which shows a relatively flat metallicity profile ($\sim-0.02$~dex/kpc 
in the range $0.2<r<0.6$~deg, or over $2.5$~kpc).

\end{itemize}

The strong evidence of tidal disruption in NGC147 suggests that these two dEs 
have experienced
markedly different evolutionary histories. NGC147 shows a  
shallower surface brightness profile and metallicity gradient than NGC185, which is
likely to result from environmental effects, rather than internal evolution
(e.g., rotation).  Our observations will be crucial for constraining models of 
the past histories of these closest dEs.

%________________________________________________________________

\section*{Acknowledgments}

The authors thank the PAndAS collaboration and the referee for useful comments.
DC, AMNF and EJB acknowledge support from an STFC Consolidated Grant.
DC wishes to kindly thank the hospitality of the Mullard Space Science
Laboratory, University College of London, where part
of this work has been carried out. Based on observations obtained with
MegaPrime/MegaCam, a joint project of CFHT and CEA/DAPNIA,
at the Canada-France-Hawaii Telescope (CFHT) which is operated
by the National Research Council (NRC) of Canada, the Institute
National des Sciences de l'Univers of the Centre National de la
Recherche Scientifique of France and the University of Hawaii.
This research made use of SAOImage DS9, developed by Smithsonian 
Astrophysical Observatory, and of the NASA/IPAC Extragalactic Database (NED), 
which is operated by the Jet Propulsion Laboratory, California 
Institute of Technology, under contract with the National 
Aeronautics and Space Administration. 

%________________________________________________________________

\bibliographystyle{mn2e}
\bibliography{biblio.bib}

%________________________________________________________________

\label{lastpage}

%________________________________________________________________

\end{document}